\newcommand{\RIKEN}{RIKEN Nishina Center, 
  2-1 Hirosawa, Wako, Saitama 351-0198, Japan}  
\newcommand{\KEK}{High Energy Accelerator Research Organization
  (KEK), Oho 1-1, Tsukuba, Ibaraki 305-0801, Japan}
\newcommand{\OU}{Department of Physics, Osaka University, 
  Machikaneyama 1-16, Toyonaka, Osaka 560-0034, Japan}
\newcommand{\RCNP}{Research Center for Nuclear Physics (RCNP), 
  Osaka University, Ibaraki, Osaka 567-0047, Japan} 

\newcommand{\TITech}{Department of Physics, Tokyo Institute of
  Technology, 2-12-1 Oh-okayama, Meguro-ku, Tokyo 152-8551, Japan}
\newcommand{\JAEA}{Japan Atomic Energy Agency, Tokai, Ibaraki
  319-1195, Japan}

\newcommand{\OAC}{Research Center for Physics and Mathematics, Osaka
  Electro-Communication University, 18-8 Hatsucho, Neyagawa, Osaka
  572-8530, Japan} 
\newcommand{\AIST}{National Institute of Advanced Industrial Science 
  and Technology (AIST), Tsukuba Central 2, 1-1-1 Umezono,
  Tsukuba,Ibaraki 305-8568, Japan}
\newcommand{\TSUKUBA}{Institute of Physics, University of Tsukuba,
  Tsukuba, Ibaraki 305-8571, Japan} 
\newcommand{\Chicago}{Enrico Fermi Institute, University of Chicago,
  5635 S. Ellis Ave., Chicago, IL 60637, USA}

 \documentclass[preprint,showpacs,amsmath,amssymb
  ,superscriptaddress]{revtex4}
\usepackage{graphicx}% Include figure files
\usepackage{dcolumn}% Align table columns on decimal point
\usepackage{bm}% bold math
\begin{document}

%%%%%%%%%%%%%%%%%%%%%%%%%%%%%%%%%%%%%%%%%%%%%%%%%%%%%%%%%%%%%%%%%%%%70
%% title 
%%%%%%%%%%%%%%%%%%%%%%%%%%%%%%%%%%%%%%%%%%%%%%%%%%%%%%%%%%%%%%%%%%%%70
%
\title{Beta-delayed neutron and gamma-ray spectroscopy of 
  $^{17}$C utilizing spin-polarized $^{17}$B} 
%
%%%%%%%%%%%%%%%%%%%%%%%%%%%%%%%%%%%%%%%%%%%%%%%%%%%%%%%%%%%%%%%%%%%%70
%% authors
%%%%%%%%%%%%%%%%%%%%%%%%%%%%%%%%%%%%%%%%%%%%%%%%%%%%%%%%%%%%%%%%%%%%70

%------------------------
\author{H.~Ueno}
\email{ueno@riken.jp}
\affiliation{\RIKEN}
%------------------------
\author{H.~Miyatake}
\affiliation{\KEK}
%------------------------
\author{Y.~Yamamoto}
\affiliation{\OU}
%------------------------
\author{S.~Tanimoto}
\affiliation{\OU}
%------------------------
\author{T.~Shimoda}
\affiliation{\OU}
%------------------------
\author{N.~Aoi}
\affiliation{\RCNP}
%------------------------
\author{K.~Asahi}
\affiliation{\TITech}
%------------------------
\author{E.~Ideguchi}	
\affiliation{\RCNP}
%------------------------
\author{M.~Ishihara}
\affiliation{\RIKEN}
%------------------------
\author{H.~Izumi}
\affiliation{\OU}
%------------------------
\author{T.~Kishida}	
\affiliation{\RIKEN}
%------------------------
\author{T.~Kubo}
\affiliation{\RIKEN}
%------------------------
\author{S.~Mitsuoka}
\affiliation{\JAEA}
%------------------------
\author{Y.~Mizoi}	
\affiliation{\OAC}
%------------------------
\author{M.~Notani}
\affiliation{\Chicago}
%------------------------
\author{H.~Ogawa}
\altaffiliation[Present Address: ]{\AIST}
\affiliation{\TITech}
%------------------------
\author{A.~Ozawa}	
\affiliation{\TSUKUBA}
%------------------------
\author{M.~Sasaki}	
\affiliation{\OU}
%------------------------
\author{T.~Shirakura}	
\affiliation{\OU}
%------------------------
\author{N.~Takahashi}	
\affiliation{\OU}
%------------------------
\author{K.~Yoneda}
\affiliation{\RIKEN}

%%%%%%%%%%%%%%%%%%%%%%%%%%%%%%%%%%%%%%%%%%%%%%%%%%%%%%%%%%%%%%%%%%%%70
%% date
%%%%%%%%%%%%%%%%%%%%%%%%%%%%%%%%%%%%%%%%%%%%%%%%%%%%%%%%%%%%%%%%%%%%70

\date{\today}% It is always \today, today,

%%%%%%%%%%%%%%%%%%%%%%%%%%%%%%%%%%%%%%%%%%%%%%%%%%%%%%%%%%%%%%%%%%%%70
%% abstract
%%%%%%%%%%%%%%%%%%%%%%%%%%%%%%%%%%%%%%%%%%%%%%%%%%%%%%%%%%%%%%%%%%%%70

\begin{abstract}
Excited states in $^{17}$C were investigated through the measurement
of $\beta$-delayed neutrons and $\gamma$ rays emitted in the $\beta$
decay of $^{17}$B. 
In the measurement, three negative-parity states and two
inconclusive states, were identified in $^{17}$C above the neutron
threshold energy, and seven $\gamma$ lines were identified in
a $\beta$-delayed multiple neutron emission of the $^{17}$B $\beta$
decay. From these transitions, the $\beta$-decay scheme of
$^{17}$B was determined. 
In particular, a de-excitation 1766-keV
$\gamma$ line from the first excited state of $^{16}$C was observed in
coincidence with the emitted $\beta$-delayed neutrons, and this
changes the previously reported $\beta$-decay scheme of $^{17}$B and
level structure of $^{17}$C. In the present work, the $\beta$-NMR
technique is combined with the $\beta$-delayed particle measurements
using a fragmentation-induced spin-polarized $^{17}$B beam. This new
scheme allows us to determine the spin parity of $\beta$-decay
feeding excited states based on the difference in the discrete
$\beta$-decay asymmetry parameters, provided the states are connected
through the Gamow-Teller transition. In this work, 
$I^{\pi}=1/2^{-}$, $3/2^{-}$, and ($5/2^{-}$) are assigned to the
observed states at $E_{\rm x}=$ 2.71(2), 3.93(2), and 4.05(2) MeV in
$^{17}$C, respectively. 
\end{abstract}

%%%%%%%%%%%%%%%%%%%%%%%%%%%%%%%%%%%%%%%%%%%%%%%%%%%%%%%%%%%%%%%%%%%%70
%% PACS & Keywords
%%%%%%%%%%%%%%%%%%%%%%%%%%%%%%%%%%%%%%%%%%%%%%%%%%%%%%%%%%%%%%%%%%%%70

%% PACS numbers may be entered using the \verb+\pacs{#1}+ command.
%                             % Classification Scheme.

\maketitle
%%%%%%%%%%%%%%%%%%%%%%%%%%%%%%%%%%%%%%%%%%%%%%%%%%%%%%%%%%%%%%%%%%%%70
%% Main body
%%%%%%%%%%%%%%%%%%%%%%%%%%%%%%%%%%%%%%%%%%%%%%%%%%%%%%%%%%%%%%%%%%%%70

%%%%%%%%%%%%%%%%%%%%%%%%%%%%%%%%%%%%%%%%%%%%%%%%%%%%%%%%%%%%%%%%%%%%70
\section{Introduction}
%%%%%%%%%%%%%%%%%%%%%%%%%%%%%%%%%%%%%%%%%%%%%%%%%%%%%%%%%%%%%%%%%%%%70

% ----------
Neutron-rich carbon isotopes are attracting because of their
anomalous level structures. It has been experimentally shown that
none of the odd mass neutron-rich carbon isotopes, $^{15-19}$C, have
the ground-state (GS) spin parities of $I^{\pi}=$ $5/2^{+}$, despite
the $d_{{5/2}}$ valence neutron expected from a naive shell model. 
% ----------
In $^{15}$C, having the neutron number $N=9$, the inversion
of single-particle levels between $d_{5/2}$ and
$s_{1/2}$ is suggested from the GS spin parity of $^{15}$C, 
$I^{\pi}_{\rm GS} =$
${1/2}^{+}$~\cite{TAL60,SUZ94_S12N9}. 
Intriguingly, the further neutron-rich nucleus $^{17}$C with $N=11$,
has shown to have $I^{\pi}_{\rm GS}=$ ${3/2}^{+}$ . This
$I^{\pi}_{\rm GS}$ assignment was performed through the study of GS
properties based on the direct
reaction~\cite{BAU98,BAZ98,SAU00,MAD01,SAT08}, $\beta$-delayed
neutron spectroscopy of the $\beta$ decay of $^{17}$C~\cite{SCH95}, 
and magnetic moment of $^{17}$C~\cite{OGW02_17C}. In the extremely
neutron-rich nucleus $^{19}$C with $N=13$, the $I^{\pi}_{\rm GS}$
value again becomes ${1/2}^{+}$; this was confirmed and discussed in
connection with the formation of a neutron 
halo~\cite{NAK99,BAZ95,MAD01}. 
%% --------------

%% -------
%% ・intro: p-sd orbitの中性子過剰核側での系統的変化を見る上では、
%% negative parity stateの共同を調べるのが良い・・・、とか  
In these neutron-rich carbon isotopes, unlike the valence protons,
which occupy the $p$ shell, the valence neutrons occupy the $sd$
shell, where the $p$-$sd$ cross-shell interactions characteristically play an
important role.
To gain an  understanding of these intriguing properties of
$p$-$sd$ neutron-rich carbon isotopes, it is important to investigate
not only the GS but also the structure of the excited states,
because those negative-parity
states can be described such that one particle in the the
$p$ shell of the low-lying positive parity states is excited to
the $sd$ shell, and thus, their energy differences from the low-lying
positive-parity states are expected to reflect directly such the
effective interactions. 

%% -------
With regard to
the excited states of $^{17}$C, the existence of three low-lying
positive-parity states has been proposed at $E_{\rm x} \sim$ 210,
295, and 330~keV below the one-neutron threshold energy $S_{\rm n}=$
0.729(18)~MeV~\cite{AU97}. An excited state observed at $E_{\rm x}=$ 
295(20)~keV~\cite{NOL77} in the
$^{48}$Ca($^{18}$O,~$^{17}$C)$^{49}$Ti reaction was again observed at 
$E_{\rm x}=$  295(10) for the same reaction, carried out at a
slightly higher beam energy~\cite{FIF82}. However, the 295-keV
state was interpreted to be a background (BG) event during the
in-beam $\gamma$-ray spectroscopy performed with a $^{17}$C
beam~\cite{KAN05}. Two more excited states were observed at  
${E_{\rm x}}=$ 207(15)~keV and 329(5)~keV in the two-step
fragmentation reaction~\cite{STA04}. Two corresponding energy levels
were observed, i.e., $E_{\rm x}=$ 210(4)~keV and 331(6)~keV, in the
$p$($^{19}$C,~$^{17}$C$\gamma$) reaction, for which $I^{\pi}$ values
were assigned as $1/2^{+}$ and $5/2^{+}$,
respectively~\cite{ELE05}. The same $I^{\pi}$ assignments were
identified in the one-neutron removal reaction of $^{18}$C from a
proton target, wherein a coupled-channel analysis was performed for
the corresponding levels $E_{\rm x}=$ 210~keV and
330~keV~\cite{KON09_17C}, and in the lifetime measurement, in which
case the M1 transition strengths were discussed for the observed
212(8)- and 333(10)-keV states~\cite{SUZ08_17C}. For high-lying
states, thirteen positive-parity states, including the $E_{\rm x}=$
310(30)~keV level, have been observed up to $E_{\rm x}=$ 16.3~MeV in
the study of the $^{14}$C($^{12}$C,~$^{9}$C)$^{17}$C
reaction~\cite{BOH04}.

Even then, only a few studies have been conducted on the
negative-parity states in $^{17}$C. For studying negative-parity
states in $^{17}$C, it is useful to perform the $\beta$-decay study
of $^{17}$B. In the light mass region of neutron-rich nuclei, the
parity of $sd$-valence neutrons differs from that of $p$-shell
valence protons, whereby $\beta$-decay allowed transitions feed
negative-parity states~\cite{MK3}. Hence, a $\beta$-decay study is
useful for studying neutron-rich nuclei located away from the
stability line, owing to the large $Q_{\beta}$ windows. Over the past
several years, a number of such studies have been performed on the
structure of light-mass neutron-rich nuclei through the
time-of-flight (TOF) measurement of emitted $\beta$-delayed
neutrons~\cite{HAR99_15B,SCH94,SCH95,RAI96_17Bbn,AOI02}.

In the present work, we performed the spectroscopic study of
{$\beta$}-delayed neutrons and {$\gamma$} rays in the {$\beta$} decay
of $^{17}$B in order to investigate the level structure of $^{17}$C
(hereafter, $\beta$-delayed neutron(s) and $\beta$-delayed $\gamma$
ray(s) are denoted as $\beta$-n and $\beta$-$\gamma$,
respectively). The $\beta$-n measurement of the $^{17}$B $\beta$
decay has thus far been reported in Ref.~\cite{RAI96_17Bbn}, and in this
study, several $\beta$-decay transitions to the excited states in
$^{17}$C above the neutron threshold energy, which were followed by
neutron emission, were observed. However, in the construction of the
decay scheme, all transitions observed in a one-neutron (1n) emission
channel were assumed to be directly connected to the GS of
$^{16}$C. In order to identify the final states in $^{16}$C
subsequent to the $\beta$-n emissions, we also conducted $\gamma$-ray
measurements in coincidence with the $\beta$ rays and
neutrons. Moreover, it should be noted that this measurement was
combined with a technique of fragmentation-induced
spin-polarization~\cite{ASAHI,OKUNO}. Thus, the GS of $^{17}$B, as an
initial state of $\beta$ decay, was spin-polarized. The angular
distribution of the $\beta$ decay through the Gamow-Teller (GT)
transition from a spin-polarized nucleus is known to show anisotropy
with respect to the polarization axis, and it is characterized by
$A_{\beta} P$, where $A_{\beta}$ denotes the asymmetry parameter of
the corresponding GT transition given by the $I^{\pi}$ value of the
initial and final states and $P$ denotes the polarization of the
parent nucleus. Since $P$ is common to all the transitions, the final
state $I^{\pi}$ can be assigned such that $A_{\beta}$ becomes
proportional to the experimentally determined  $A_{\beta} P$ values
when the initial state $I^{\pi}$ is known. For the first time, by
using this new method, the $I^{\pi}$ values of the excited states in
$^{15}$C have been successfully assigned with the spin-polarized
$^{15}$B beam~\cite{MIYA03}. Further, by taking the advantage of
highly spin-polarized Li and Na beams, the nuclear structures of
$^{11}$Be~\cite{HIR05} and $^{28}$Mg~\cite{SHI12} have been
studied through the spin-parity assignment of the excited states.

%%%%%%%%%%%%%%%%%%%%%%%%%%%%%%%%%%%%%%%%%%%%%%%%%%%%%%%%%%%%%%%%%%%%70
%%
%%
\section{Experimental procedure}
%%
%%
%%%%%%%%%%%%%%%%%%%%%%%%%%%%%%%%%%%%%%%%%%%%%%%%%%%%%%%%%%%%%%%%%%%%70

%%======================================================================
%%
\subsection{Production of spin-polarized $^{17}$B  beam}
%%
%%======================================================================
A spin-polarized $^{17}$B beam was produced using the same procedure
as described in Ref.~\cite{HU17B}, in which the fragmentation-induced
spin polarization technique was adopted~\cite{ASAHI,OKUNO}. A beam of
$^{17}$B  was obtained from the fragmentation of a $^{22}$Ne
projectile with an energy of $E/A=$ 110~MeV and a current of 90
particle-nA incident on a $^{93}$Nb target having a thickness of
1.07~g/cm$^2$. In order to obtain a spin-polarized $^{17}$B beam, the
emission angles and outgoing momenta of the $^{17}$B fragments were
suitably selected~\cite{ASAHI,OKUNO}. Thus, fragments emitted within
radial angles $\theta_{\rm{L}} = 1.5^{\circ}-5.0^{\circ}$ and
azimuthal angles $|\phi_{\rm L}| \leq 2.0^{\circ}$ along with the
primary beam were accepted by the RIKEN projectile-fragment separator
RIPS~\cite{RIPS}, using a beam swinger installed upstream of the
target. A range of the momentum values, i.e., $p=$ 7.11 to 7.55
GeV/$c$, was selected with the help of a slit at the intermediate
momentum-dispersive focal plane. This momentum range corresponds to
the range 1.01$p_{0}$--1.07$p_{0}$, where $p_{0} = 7.03$~GeV/$c$ is
the fragment momentum corresponding to the projectile velocity. The
isotope separation was given by the combined analyses of the magnetic
rigidity and the momentum loss in a wedge-shaped degrader~\cite{RIPS}
with a median thickness of 1638~mg/cm$^2$ and a slope angle of
8.67~mrad. 

The spin-polarized and isotope-separated $^{17}$B fragments were then
introduced into an apparatus located at the final focus of RIPS, for
the $\beta$-delayed particle measurement. They were implanted in a Pt
stopper located at the center of the apparatus, which consisted of a
stack of four 100-$\mu$m-thick Pt plates. The beam implantation was
confirmed with plastic scintillators placed upstream and downstream
of the Pt stopper. The the upstream counter was used to distinguish
the $Z>4$ beam particles from contaminating tritons. Under the
conditions described above, RIPS provided a $^{17}$B beam  with a
purity of $\sim$100\%, not considering tritons, and an intensity of
17.3 particles per second (pps). 

%%======================================================================
%%
\subsection{Detector apparatus}
%%
%%======================================================================
We combined the $\beta$-ray detected nuclear magnetic resonance
($\beta$-NMR) technique~\cite{SUG66} with the 
$\beta$-delayed particle emission measurement. A static magnetic
field, $B_{0}=50.0$ mT, whose inhomogeneity was  
${\Delta}{B_{0}}/{B_{0}} \leq 10^{-2}$, was applied to the Pt stopper
with a Helmholtz-type air coil in order to preserve the spin
polarization of $^{17}$B. The effective diameter of the coil was
$\phi=$ 250~mm. $\beta$-Rays emitted from the implanted fragments
were detected with $\beta$-ray telescopes, consisting of two plastic
scintillators, located above and below the Pt stopper. To exclude BG
events, such as those involving cosmic muons, coincidences with
signals from the other side of the $\beta$-ray telescope were checked
as soon as a $\beta$-ray was detected.  

\begin{figure}[tbh]
\begin{center}
\includegraphics[width=.80\textwidth]{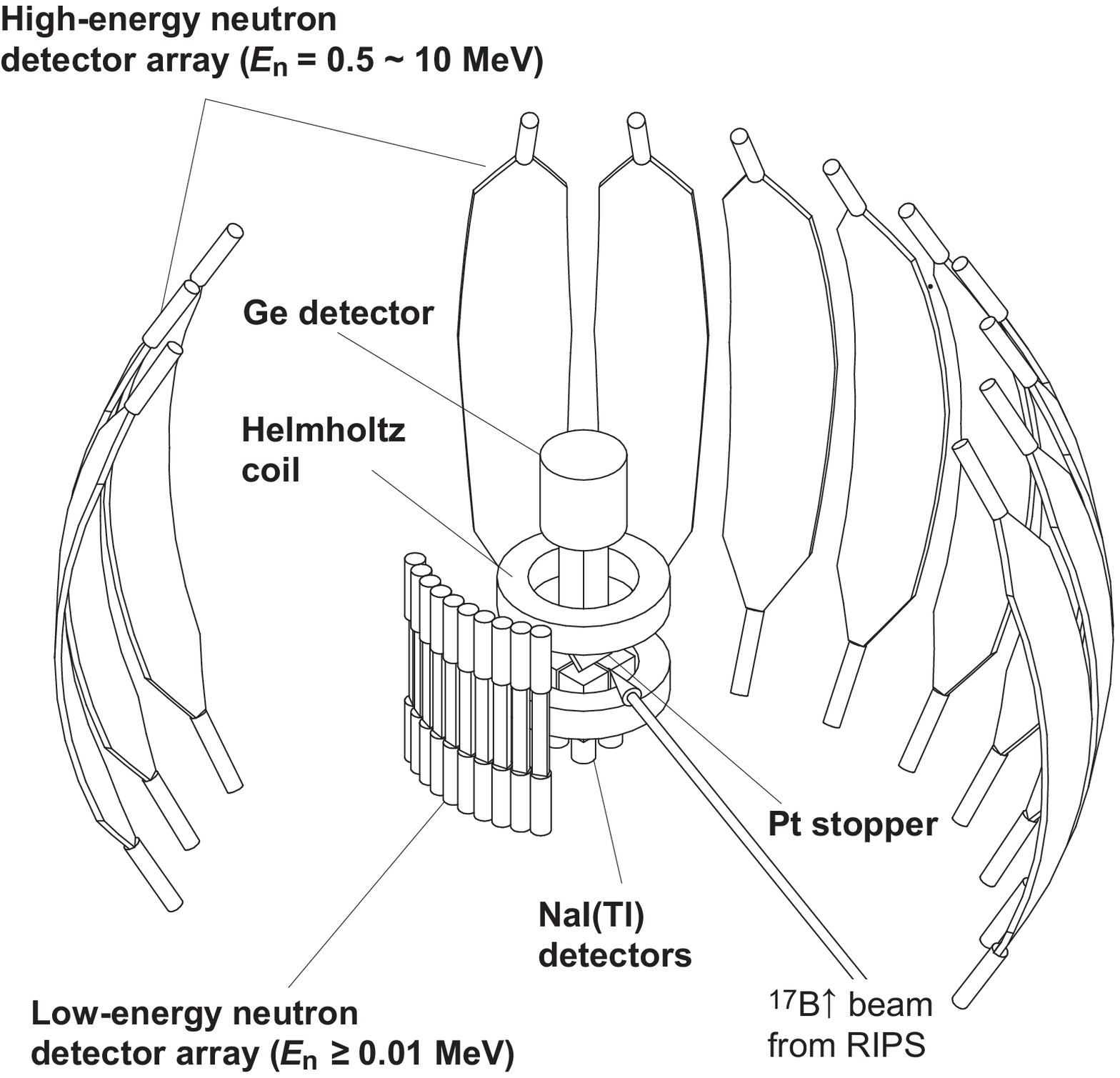}
\caption{Arrangement of the high-energy and low-energy neutron
  detector arrays.
  \label{F:NEUT-ARRAY}}
  \end{center}
\end{figure}

%% ------------------------------------------------------
%% Neutron detectors
%% ------------------------------------------------------
\label{S:NeutEff}
The $\beta$-n emitted from the implanted $^{17}$B were measured with
a high-energy neutron detector array~\cite{AOI02,MIYA03}, consisting
of a set of twelve plastic scintillators, whose shapes were curved in
the vertical direction with a 150-cm radius, 160-cm arc length, and
40-cm latitudinal width in the median plane. The achieved
electron-equivalent threshold energy, $E_{\rm th}=$ 3.4(27)~keVee, of
the neutron detector array, enabled the detection of low-energy
neutrons at $E_{\rm n} \sim 0.5$~MeV, and an efficiency of 5.6\% for
$\sim$1~MeV neutrons was achieved as a result. A high-energy neutron
detector array was placed 1.5~m away from the Pt stopper, as shown in
Fig.~\ref{F:NEUT-ARRAY}, covering the solid angles 
$\Omega_{\rm{n}} = 0.21{\times}4{\pi}$ sr. In this configuration,
neutrons emitted vertical to the $\beta$-ray direction were detected,
which minimized the neutron-energy broadening due to $\beta$-ray
recoil effects. The neutron energies were determined using the TOF
method, wherein the $\beta$-ray signal was used as the start
pulse. The signal was read out from photomultiplier tubes (PMTs)
attached at each end, and both the read out signals were used to
determine the correct TOF by calculating the mean time. Neutrons
within the energy range $E_{\rm n} = 0.5 \sim 10$~MeV were analyzed
with this high-energy neutron detector array. 

%%------------------------------------------------------------------70
%% FIG1 should be planced around here
%%------------------------------------------------------------------70
%  \begin{figure}[tbh]
%  \begin{center}
%  \includegraphics[width=.80\textwidth]{FIG1_UENO_NARRAY.eps}
%  \caption{Arrangement of the high-energy and low-energy neutron
%    detector arrays.
%    \label{F:NEUT-ARRAY}}
%    \end{center}
%  \end{figure}
%%------------------------------------------------------------------70

The detection efficiency of the neutron detector array was determined
using $^{15}$B and $^{17}$N beams, whose $\beta$-decay branches
associated with $\beta$-n emission are
known~\cite{OWM76_17Nbn,HAR99_15B}. Using these calibration data, the
efficiency was determined as a function of the energy $E_{\rm n}$,
based on the simulation code given in Ref.~\cite{CEC79}, where the
reduction of photo-propagation  in the plastic scintillator was taken
into account. Details pertaining to the high-energy neutron detector
array are given in Ref.~\cite{MIYA03}. For the measurement of
neutrons having an energy down to $E_{\rm n} \gtrsim 0.01$~MeV, a
low-energy neutron detector array~\cite{AOI02} covering the solid
angle $\Omega_{\rm n}=$ 0.037{$\times$}4{$\pi$}~sr was placed on a
concentric circle to achieve a distance of 0.5~m from the Pt stopper,
which consisted of a set of ten
45~mm~{$\times$}~25~mm~{$\times$}~300~mm plastic scintillators. The
signal was read out from the PMTs attached at each end. The threshold
energy was set to 2~keVee using the Compton edge of $^{137}$Cs. 

\begin{figure}[bth]
\begin{center}
 \includegraphics[width=.80\textwidth]{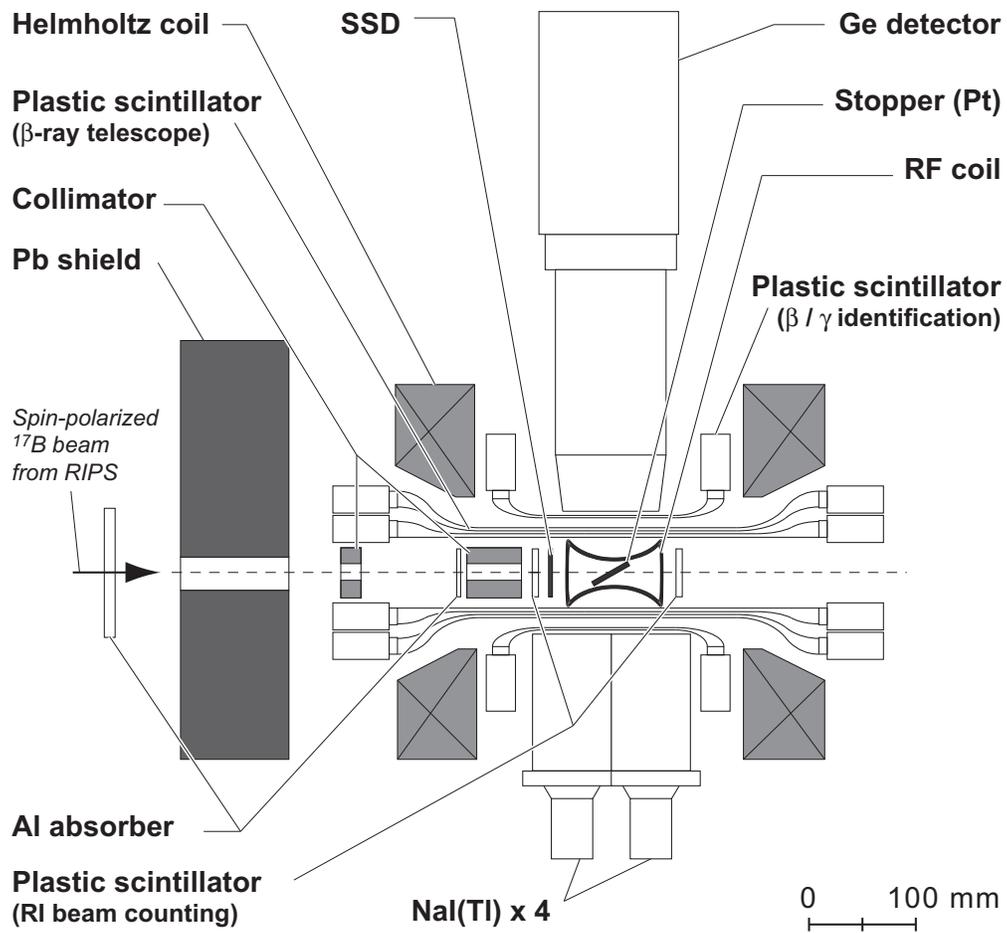}
 \caption{Schematic layout of the setup around a Pt stopper, showing
  NaI(Tl) and Ge $\gamma$-ray detectors and a $\beta$-NMR system. 
  \label{F:NMR}}
  \end{center}
\end{figure}

%% ------------------------------------------------------
%% Gamma-ray detectors 
%% ------------------------------------------------------
In addition to the neutron counters, a 50~mm$^{\phi}~\times$~70~mm
Clover Ge detector~\cite{JON95} and a set of four
80~mm~$\times$~80~mm~$\times$~152~mm NaI(Tl) detectors were placed
above and below the $\beta$-ray telescope, as shown in
Fig.~\ref{F:NMR}. The PMTs housed in the NaI(Tl) detectors are a type
of fine-mesh dynodes, which can be operated under strong magnetic
fields with strengths of over 1~T~\cite{HAMA6614}. 
%% ------------------------------------------------------
%% Gamma-ray detectors for beta-ray energy measurement
%% ------------------------------------------------------
\label{Sec:NaIforBeta}
In front of each Ge and NaI(Tl) detector, 2-mm-thick plastic
scintillators were placed in order to distinguish $\gamma$ and
$\beta$ rays. With the help of these plastic scintillators, NaI(Tl)
detectors were used to measure not only $\gamma$-ray energies but
also the $\beta$-ray total energy, $E_{\beta}$, for up to 30~MeV,
which covers a $Q_{\beta}(^{17}{\rm B})$ of
22.68(14)~MeV~\cite{AU97}. The $E_{\beta}$ data were calibrated for 
a sufficiently wide range using $^{17}$N and $^{15}$B, whose
$Q_{\beta}$ values are 8.680(15) and 19.094(22)~MeV,
respectively~\cite{AU97}. 

%%------------------------------------------------------------------70
%% FIG2 should be planced around here
%%------------------------------------------------------------------70
%  \begin{figure}[bth]
%  \begin{center}
%   \includegraphics[width=.80\textwidth]{FIG2_UENO_STP.eps}
%   \caption{Schematic layout of the setup around a Pt stopper,
%   showing NaI(Tl) and Ge $\gamma$-ray detectors and a $\beta$-NMR
%   system. 
%    \label{F:NMR}}
%    \end{center}
%  \end{figure}
%%------------------------------------------------------------------70

%%==================================================================70
%%
\subsection{Block diagram}
%%
%%==================================================================70

The conventional block diagram for $\beta$-NMR measurements shows a
beam being pulsed with a beam bombardment period of $T_{\rm B}$. At
the beginning of the beam-off period in these measurements, the
oscillating magnetic field $B_1$ is applied for a duration of 
$T_{\rm R}$, with its frequency swept over a Larmor frequency in
order to reverse the direction of the spin polarization by means of
the adiabatic fast passage (AFP) NMR method~\cite{ABRAGAM}. Then,
$\beta$ rays are counted during the following period, 
$T_{\rm C}$. This unit cycle, $T_{\rm B} + T_{\rm R} + T_{\rm C}$, is
repeated many times until sufficient statistics have been
accumulated. Note that the time length of the unit cycle is generally
set to  $2/\lambda$, where $\lambda$ is the decay constant of the
nuclei of interest. Through the execution of this sequence, a maximum
figure of merit given by $Y_{\beta} \times P^{2}$ is achieved, where
$Y_{\beta}$ is the $\beta$-ray counting rate and $P$ is the nuclear
spin polarization. Ideally, $T_{\rm B}$ should be set by taking into
account the spin-lattice relaxation time of the nuclei. The
$\beta$-ray counting rate $Y_{\beta}$ is then given by  
%%------------------------------------------------------------------70
%\begin{widetext}
\begin{eqnarray}
Y_{\beta} 
= \frac{I_{\rm B} 
      e^{-\lambda T_{\rm R}}
      (1-e^{-\lambda T_{\rm B}})
      (1-e^{-\lambda T_{\rm C}})
  }{  \lambda 
      (T_{\rm B}+T_{\rm C}+T_{\rm R}) 
      (1-e^{-\lambda (T_{\rm R}+T_{\rm C})})
  },
\label{Eq:YbNormalCycle}
\end{eqnarray}
%\end{widetext}
%%------------------------------------------------------------------70
where $I_{\rm B}$ is the beam-implantation rate during the period
$T_{\rm B}$. In this sequence, however, $\sim$50\% of the beam needs
to be blocked, despite the low production yield of the $^{17}$B beam.

\begin{figure}[bth]
\begin{center}
 \includegraphics[width=.80\textwidth]{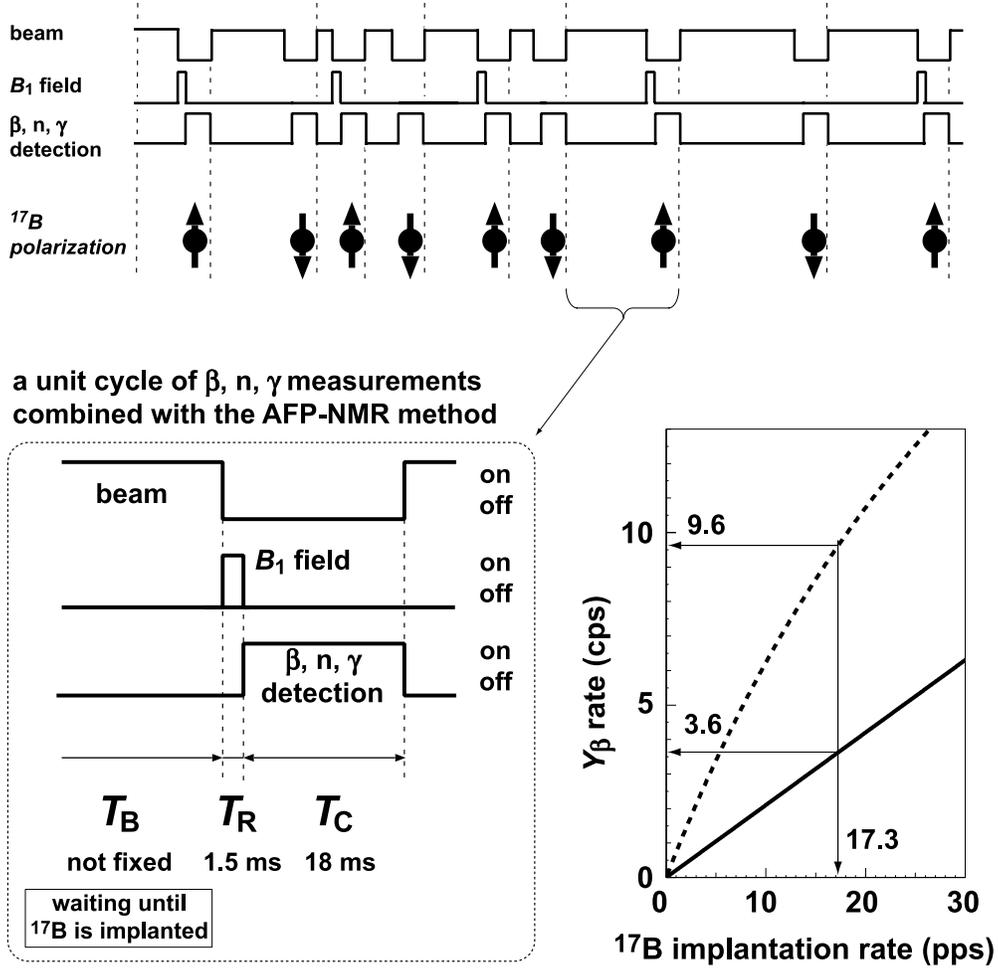}
 \caption{Block diagram for the $\beta$-delayed neutron and/or
  $\gamma$ ray measurement combined with the AFP-NMR
  technique. $T_{\rm B}$, $T_{\rm R}$, and $T_{\rm C}$ are the
  duration of the $^{17}$B beam bombardment; the application of the
  $B_{1}$ field; and the $\beta$, neutron, and $\gamma$ particle
  detection. $T_{\rm{R}}$ and $T_{\rm{C}}$ are fixed, while
  $T_{\rm B}$ is not fixed and remains open until a $^{17}$B particle
  is implanted. The $\beta$ decay rate $Y_{\beta}$, as the function
  of the $^{17}$B implantation rate, of the new {\sl beam waiting}
  mode was compared with that of the fixed beam-on/off cycle
  mode. For details, see the text.
\label{F:TIME_CHART}}
\end{center}
\end{figure}

For  more efficient measurements, the block diagram was improved in
the present study, and it is illustrated in
Fig.~\ref{F:TIME_CHART}. The beam was not pulsed periodically at
fixed durations. Instead, as soon as a $^{17}$B particle was
identified by the beam-line counters equipped with RIPS, the beam
bombardment was turned off for 19.5~ms. At the beginning of the
beam-off period, the $B_1$ field $\simeq$ 1.6~mT was applied for a
duration of $T_{\rm R}=$ 1.5~ms, and in this duration, the frequency
of $B_1$ was swept over the Larmor frequency $\nu_0$ of
$^{17}$B~\cite{HU17B} in the frequency window $\Delta \nu/\nu_0$ =
2\%. Then, the $\beta$ rays, in addition to $\gamma$ rays and
neutrons, were measured during a subsequent time period of 
$T_{\rm C}=$ 18~ms. After the beam-off period, the beam bombardment
was again turned on until the next $^{17}$B particle was
detected. The $B_1$ field was applied every two beam-off cycles, to
ensure that the direction of the spin polarization would change
alternately; this reduced the systematic error due to differences in
the efficiencies of the two $\beta$-ray telescopes and long-term
beam-profile fluctuation. The $Y_{\beta}$ yield in this 
{\sl beam-waiting} mode is given by 
%%------------------------------------------------------------------70
%\begin{widetext}
\begin{eqnarray}
Y_{\beta} 
= \frac{ I_{\rm B}
     \left(
         e^{-\lambda T_{\rm R}}-e^{-\lambda(T_{\rm R}+ T_{\rm C} )}
     \right)
  }{1+I_{\rm B}(T_{\rm R}+ T_{\rm C})}.
\label{Eq:YbBeamTrigger}
\end{eqnarray}
%\end{widetext}
%%------------------------------------------------------------------70
Based on Eqs.~(\ref{Eq:YbNormalCycle}) and (\ref{Eq:YbBeamTrigger}),
the yields of $^{17}$B $Y_{\beta}$ were calculated, for the purpose
of comparison, as a function of the intensity of a $^{17}$B beam in
the fixed beam-on/off cycle mode and the {\sl beam-waiting} mode,
where $(T_{\rm B}, T_{\rm R}, T_{\rm C}) = (7.3, 1.5, 7.3)$~ms was
assumed as a typical sequence in the former mode. As shown in the
inset of Fig.~\ref{F:TIME_CHART}, $Y_{\beta} = 9.6$ counts per second
(cps) for the {\sl beam-waiting} mode, and it is 2.6 times larger
than $Y_{\beta} = 3.6$ cps, which is observed for the fixed
beam-on/off cycle mode under an actual $^{17}$B intensity of 
$I_{\rm B}=17.3$ pps. Factually, 12.9-pps $^{17}$B particles were
implanted at the intensity of $I_{\rm B}=17.3$ pps because of the
dead time due to the 19.5-ms beam blocking period 
$T_{\rm R}+T_{\rm C}$. This effect is already considered in
Eq.~(\ref{Eq:YbBeamTrigger}).

Another important advantage of the new {\sl beam-waiting} mode is the
high S/N ratio in the $\beta$-ray measurements. Given the reported
multiple neutron emission probabilities in the $^{17}$B $\beta$
decay~\cite{DUF88}, 3.6 $\beta$ rays are emitted on an average in
the $\beta$-decay chain initiated by one $^{17}{\rm B}$ $\beta$
decay. Since all decay-chain nuclei have significantly longer
lifetimes than $^{17}$B, the BG $\beta$ rays from these nuclei are
assumed to be detected with the same  probability inside  $T_{\rm C}$
windows. For $I_{\rm B}=17.3$ pps, the S/N ratio in the fixed
beam-on/off mode is S(3.6 cps)$/$N(16.8 cps) $=$ 0.22, whereas in the
{\sl beam-waiting} mode, the S/N ratio is 4.3 times better and has
the value S(9.6 cps)$/$N(10.5 cps) $=$ 0.92. 

%%------------------------------------------------------------------70
%% FIG3 should be planced around here
%%------------------------------------------------------------------70
%  \begin{figure}[bth]
%  \begin{center}
%   \includegraphics[width=.80\textwidth]{FIG3_UENO_BD.eps}
%   \caption{Block diagram for the $\beta$-delayed neutron and/or
%    $\gamma$ ray measurement combined with the AFP-NMR
%    technique. $T_{\rm B}$, $T_{\rm R}$, and $T_{\rm C}$ are the
%    duration of the $^{17}$B beam bombardment; the application of
%    the $B_{1}$ field; and the $\beta$, neutron, and $\gamma$
%    particle detection. $T_{\rm{R}}$ and $T_{\rm{C}}$ are fixed,
%    while $T_{\rm B}$ is not fixed and remains open until a $^{17}$B
%    particle is implanted. The $\beta$ decay rate $Y_{\beta}$, as
%    the function of the $^{17}$B implantation rate, of the new 
%    {\sl beam waiting}
%    mode was compared with that of the fixed beam-on/off cycle
%    mode. For details, see the text.
%  \label{F:TIME_CHART}}
%  \end{center}
%  \end{figure}
%%------------------------------------------------------------------70

%%==================================================================70
%%
\subsection{Principle of the spin parity assignment}
%%
%%==================================================================70
\label{S:SP_ASSIGN}
The angular distribution function $W(\theta)$ for the  $\beta$ rays
emitted from $^{17}$B with the spin polarization $P$ is given by
%%------------------------------------------------------------------70
\begin{eqnarray}
W(\theta ) \propto 1 + (v/c)A_{\beta}P \cos \theta
\end{eqnarray}
%%------------------------------------------------------------------70
where $\theta$ denotes the angle between the direction of the $\beta$
emission and the axis of the nuclear polarization, $v$ and $c$ are
the velocities of the $\beta$ particles and light, respectively, and
$A_{\beta}$ is the asymmetry parameter. For simplicity, we use the
approximation $v/c \simeq 1$, since only a high-energy portion of the
$\beta$ spectrum is included in the analysis. Then, the asymmetry
$A_{\beta}P$ of the $\beta$-decay transition feeding neutron
emissions is given by 
%%------------------------------------------------------------------70
\begin{eqnarray}
A_{\beta}P = \frac{{\sqrt \rho  - 1}}{{\sqrt \rho  + 1}}.
\label{EQ:AP}
\end{eqnarray}
%%------------------------------------------------------------------70
Here, using the measured $\beta$-n spectra recorded with the
identification of the signal-detected $\beta$-ray telescope, the
double ratio $\rho$ in Eq.~(\ref{EQ:AP}) is given by
%---------------------------------------------------- 
\begin{eqnarray}
\rho = \left( \frac{N_{\rm U}}{N_{\rm D}} \right) \bigg/
    \left( \frac{N_{\rm U}^{*}}{N_{\rm D}^{*}} \right),
\end{eqnarray}
%%------------------------------------------------------------------70
where ${N_{\rm U,D}}$ are the relevant peak counting rates in the
neutron TOF spectra measured in coincidence with signals from the
$\beta$-ray telescopes located above (denoted by U) and below
(denoted by D) the Pt stopper and ${N_{\rm U,D}^{*}}$ are those
obtained with the resonant $B_{1}$-field  application. With the
obtained $A_{\beta}P$ values, asymmetry parameters $A_{\beta}$ can be
determined provided the polarization $P$ is known.

In the case that the spin $I_{\rm i}$ state decays through the pure
GT $\beta^{\mp}$-decay transitions associated with the spin change
$\Delta I$, the $\beta$-decay asymmetry parameter $A_{\beta}$ value 
is given by
%%------------------------------------------------------------------70
\begin{eqnarray}
A_{\beta} = 
\left\{
  \begin{array}{@{\,}ll}
  \mp 1             & \mbox{($\Delta I = -1$)}\\
  \mp 1/(I_{\rm i}+1) & \mbox{($\Delta I =  \ \ 0$)}\\
  \pm I_{\rm i}/(I_{\rm i}+1) & \mbox{($\Delta I = +1$)}
  \end{array}
\right. . \label{EQ:AofGT}
\end{eqnarray}
%%------------------------------------------------------------------70
By comparing the determined $A_{\beta}$ values to the value given in
Eq.~(\ref{EQ:AofGT}), the final spin parity $I^{\pi}_{\rm f}$ can be
determined in the case of a pure GT $\beta$-decay.

In order to apply this method, it is necessary to know  
$I^{\pi}_{\rm i}$ and one $I^{\pi}_{\rm f}$ value from among the GT
$\beta$-decay transitions for the determination of the spin
polarization $P$. In the present experiment, $I^{\pi}_{\rm i}=$
$3/2^{-}$ is known for the GS of $^{17}$B~\cite{HU17B}, whereas none
of the final state $I^{\pi}_{\rm f}$ values are known. Therefore,
similar to the $I^{\pi}$ assignment for $^{15}$C~\cite{MIYA03}, the
following method was performed.

For a given set of GT transitions to the level $I_{\rm f}^{j}$
($j$=1, 2, $\cdots$, $n$), there are 3$^{n}$ combinations of the
possible $A_{\beta}$ values. For each allocated ${A_{\beta}}_{j}$
value, the polarization is evaluated as 
$P_{j} = {(A_{\beta} P)}_{j} / {A_{\beta}}_{j}$, 
where ${A_{\beta}P}_{j}$ is the measured asymmetry. For the proper
combination of $A_{\beta}$ values, i.e., for the correct spin
assignments for all relevant final states, the evaluated $P_{j}$
values need to be consistent with each other. It is therefore
expected that for improper combinations of $A_{\beta}$ values, the
expected value of $P$ will have systematic errors and the variance of
$P$ will increase. In other words, a correct assignment should reveal
the least variance of $P$. This is the guiding principle for
determining $I^{\pi}_{\rm f}$ values simultaneously. For the $i$th
set of ${A_{\beta}}_{j}$ values, the mean spin polarization
$\bar{P_{i}}$ is calculated as  
%%------------------------------------------------------------------70
\begin{eqnarray}
\bar{P_{i}} = 
\frac{\sum_{j}P_{j} w_{j}}{\sum_{j}w_{j}}p
\label{EQ:Pmean}
\end{eqnarray}
%%------------------------------------------------------------------70
where $w_j$ is the statistical weight factor for each $P_j$ value,
%---------------------------------------------------- 
\begin{eqnarray}
w_{j} = 
\frac{{A_{\beta}}_{j}^{2}}{\sigma^{2}_{(A_{\beta}P)_{j}}}
\end{eqnarray}
%%------------------------------------------------------------------70
with the experimental error $\sigma_{(A_{\beta}P)_{j}}$ for
$(A_{\beta}P)_{j}$. By definition, the reduced $\chi_{\nu}^{2}$ for
the $i$th set is given as
%---------------------------------------------------- 
\begin{eqnarray}
\chi_{\nu}^{2} (\bar{P_{i}}) = 
\frac{1}{\nu}
\sum_{j} (P_j - \bar{P_i} )^{2}w_{j} \ \ \ \ (i=1,\ 2,{\cdots},\ 3^{n})
\label{EQ:chisq}
\end{eqnarray}
%%------------------------------------------------------------------70
where $\nu$ denotes the degree of freedom ($\nu=$ $n-1$). The
above-mentioned guiding principle is that the most probable set of
${A_{\beta}}_{j}$ values should yield the least $\chi_{\nu}^{2}$ among the
3$^{n}$ values for all possible combinations of ${A_{\beta}}_{j}$
values. In such a case, $\bar{P_{i}}$ can be regarded as the
statistically expected spin polarization $P$, whose error is defined
by 
%%------------------------------------------------------------------70
\begin{eqnarray}
\sigma_{\bar{P_{i}}} = 
\frac{\sum_{j} (\bar{P_{i}}-P_{j})^2 w_{j}}{\sum_{j} w_{j}}.
\label{EQ:Pgosa}
\end{eqnarray}
%%------------------------------------------------------------------70

%%%%%%%%%%%%%%%%%%%%%%%%%%%%%%%%%%%%%%%%%%%%%%%%%%%%%%%%%%%%%%%%%%%%70
%%
%%
\section{Analysis and Results}
%%
%%
%%%%%%%%%%%%%%%%%%%%%%%%%%%%%%%%%%%%%%%%%%%%%%%%%%%%%%%%%%%%%%%%%%%%70

The $^{17}$B nucleus has the ground-state spin parity 
$I^{\pi} = 3/2^{-}$. Its $\beta$ decay, whose halflife is $t_{1/2}=$ 
5.08(5)~ms and $Q_{\beta}$ is 22.68(14)~MeV~\cite{AU97}, can be
characterized by a $\beta$-delayed multiple neutron
emission~\cite{DUF88,REE91}, for which multiplicities $M_{\rm n}$
of up to four have been reported~\cite{DUF88}. In the following
analysis, the decay property was investigated by classifying the
multiplicity of the emitted $\beta$-n. 

%%==================================================================70
%%
\subsection{0n-decay branches}
%%
%%==================================================================70

The 0n mode of the $^{17}{\rm B}$ $\beta$ decay, i.e., a $\beta$
decay not followed by $\beta$-n emission, can feed excited states in
$^{17}$C below the neutron separation energy $S_{\rm n}=$
0.729(18)~MeV or the GS. Until now, the existence of three excited 
states has been suggested at the excitation energy values 
$E_{\rm x} \sim$ 210~keV~\cite{STA04,ELE05,KON09_17C,SUZ08_17C},
295~keV~\cite{NOL77, FIF82}, and
331~keV~\cite{STA04,ELE05,KON09_17C,SUZ08_17C,BOH04}, below
$S_{\rm n} = 0.729(18)$~MeV. Investigation of these three states is a
central subject for the 0n channel. 

\subsubsection{$\gamma$-Ray energy spectra obtained from the
  $\beta$-$\gamma$ coincidence measurements} 

The observed $\gamma$-ray energy peaks below $S_{\rm n} =$
0.729(18)~MeV, measured with the Ge detector in coincidence with the
$\beta$ ray from $^{17}$B, are summarized in
Table~\ref{T:LOWLYING_GAMMA}. For a comparison, details of $\gamma$
rays observed in a BG measurement without a beam and in a
detector-calibration measurement with a $^{17}$N beam are also listed
in Table~\ref{T:LOWLYING_GAMMA}. First, the $\gamma$ rays observed in
the BG measurement, i.e., $E_{\gamma}=$ 77(2), 242(1), 352(1), and
609(2)~keV, as well as the 511-keV annihilation $\gamma$ ray, can be
excluded from the $^{17}$B $\beta$-$\gamma$ candidates. They can be
assigned to KX or the $\gamma$ rays from the lead parts of the
experimental setup, in which some amount of the U/Th decay-chain
isotopes $^{214}$Pb and $^{214}$Bi are considered to be included. The
reason that the 242(1)-keV $\gamma$ ray was not observed in the
$^{17}$N measurement is not clear. This may be attributable to the
short measurement time. Next, the $\gamma$ rays observed in the
$^{17}$N $\beta$-$\gamma$ measurement, i.e., $E_{\gamma} = 596(5)$
and 696(8), can be also excluded. They originate from the
(n,~$\gamma$) or (n,~n'$\gamma$) reactions, given that their shapes
are broadened and skewed towards higher energy, which is a typical
feature that reveals the neutron-recoil effect in Ge detectors. Here,
a 67(1)-keV peak observed in both the $^{17}$B and the $^{17}$N
measurements can be assigned to the ${\rm KX}_{\alpha 1}$ rays
originating from the Pt stopper excited by the $\beta$ rays. The
remaining $\gamma$-ray peaks at 295(2) and 331(2)~keV, as well as a
small peak at $E_{\gamma}=$ 217(2)~keV, are thus the potential
$\gamma$ rays emitted from $^{17}$C subsequent to the $^{17}$B
$\beta$ decay.

\subsubsection{Properties of the peaks at $E_{\gamma}=$ 217, 295,
  and 331~keV}

\begin{figure}[bth]
\begin{center}
 \includegraphics[width=.80\textwidth]{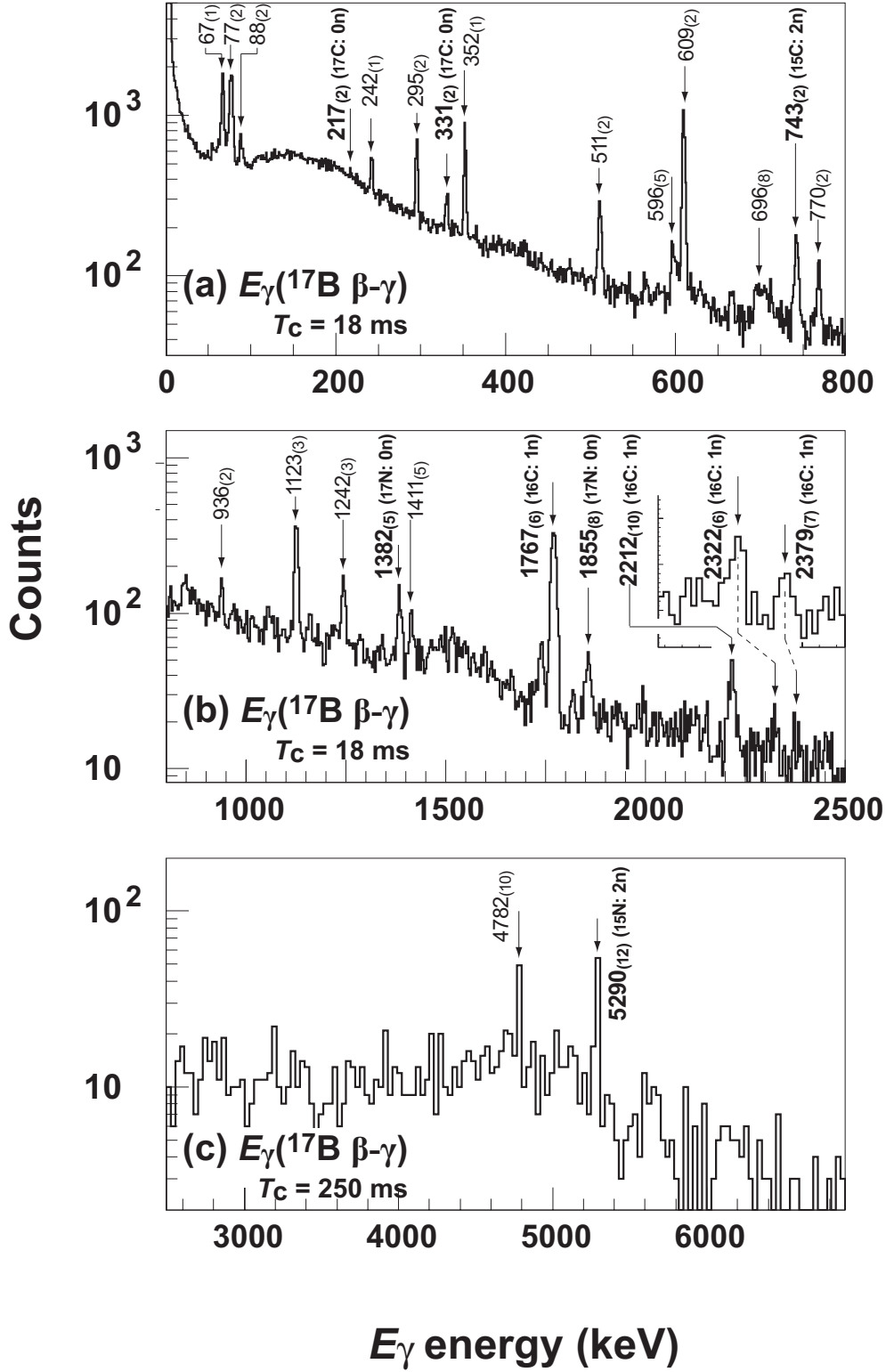}
 \caption{$\gamma$-Ray energy spectra obtained with the Ge detector in
  coincidence with the $\beta$ ray. The spectra were obtained with
  the counting periods $T_{\rm C}$ of (a) and (b) 18~ms and (c)
  250~ms. 
\label{F:GAMMA}}
\end{center}
\end{figure}

\begin{figure}[bth]
\begin{center}
 \includegraphics[width=.80\textwidth]{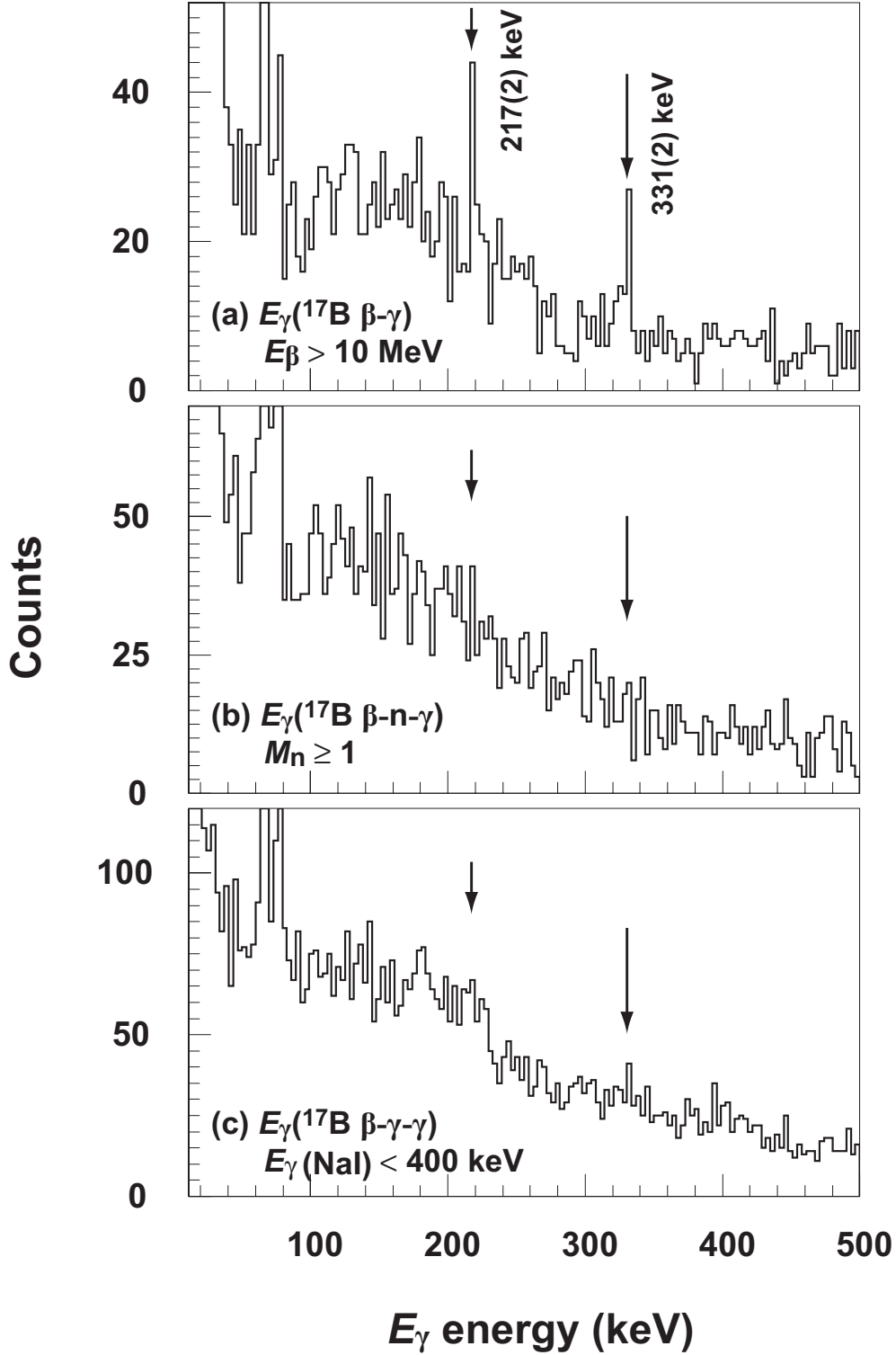}
 \caption{Obtained $\gamma$-ray energy spectra measured with the Ge
  detector by gating with (a) $\beta$-ray energies 
  $E_{\beta} > 10$~MeV, (b) neutron multiplicity $M_{\rm n} \geq$ 1,
  and (c) $\gamma$-ray energy $E_{\gamma} <$ 400~keV  measured with
  NaI(Tl) detectors.
\label{F:GAMMAgtd}}
\end{center}
\end{figure}

Properties of these $\gamma$ rays were then investigated in terms of
the total energy of the feeding $\beta$ ray, which was measured with
the set of NaI(Tl) detectors and 2-mm-thick plastic scintillators
described in Sec.~\ref{Sec:NaIforBeta}. The de-excitation $\gamma$
rays from low-lying excited states of $^{17}$C are associated with
large $\beta$-ray energies, where the maximum $\beta$-ray energies
are $E_{\beta}^{\rm max} \geq Q_{\beta}-S_{\rm n} \simeq 22$~MeV,
owing to the small $S_{\rm n}$ value. Thus, an $E_{\gamma}$ spectrum
was obtained, as shown in Fig.~\ref{F:GAMMAgtd}(a), by gating the
$E_{\gamma}$ spectrum shown in Fig.~\ref{F:GAMMA}(a) with the
$\beta$-ray energy at $E_{\beta} \ge 10$~MeV. We note that the
331(2)-keV $\gamma$ ray was still observed, while the 295(2)-keV peak
disappeared. In addition, a minor peak at $E_{\gamma}=217(2)$~keV, as
seen in Fig.~\ref{F:GAMMA}(a), is clearly observed in
Fig.~\ref{F:GAMMAgtd}(a). Although the Th decay-chain isotope
$^{228}$Ac could be a possible BG event, the 217(2)-keV $\gamma$ ray
was observed in neither the BG nor $^{17}$N measurements, suggesting
that  the 217(2)-keV peak as well as 331(2)-keV peak originate from
the $^{17}$B $\beta$ decay.

These $\gamma$ rays were further investigated in order to determine
their decay parent nuclei based on (i) the coincidence to $\beta$-n
emission and (ii) the time evolution of the $\gamma$-ray peak
counting rates. 
%% -------------------------
%% b-n-g tripple coincidence
%% -------------------------
For determining (i), a $\gamma$-ray energy spectrum obtained with the
a Ge detector  was plotted for $\beta$-n-$\gamma$ triple
coincidence events. From the resulting $E_{\gamma}$ spectrum shown in
Fig.~\ref{F:GAMMAgtd}(b), no peaks were identified within the given
statistics at $E_{\gamma} =$ 217(2) and 331(2)~keV, indicating that
these $\gamma$ rays were not associated with neutron-emission
channels in the $^{17}$B $\beta$ decay. 
%% -------------------------
%% time evolution of gamma
%% -------------------------
As described by point (ii), the time evolution of the photo-peak
counting rates was deduced to identify a parent nucleus. A time stamp
recorded with $E_{\gamma}$ was the time of a $\beta$-$\gamma$
coincidence event measured from the arrival time of a $^{17}$B
particle, which started at $T_{\rm C}$, as shown in
Fig.~\ref{F:TIME_CHART}. Thus, the time evolution in this analysis
provides the $\beta$-decay lifetime of a parent nucleus feeding the
relevant $\gamma$ emission. The obtained time spectra for 217(2),
295(2), and 331(2)~keV are shown in Figs.~\ref{F:GMLIFE}(a), (c), and
(d), respectively. Dotted lines shown in the spectra represent the
result of a least $\chi^{2}$ fitting analysis with an exponential
function having a known $^{17}$B halflife,  $t_{1/2}(\rm{^{17}B})$,
plus a constant. The time spectra of 217(2) and 331(2)~keV are
well reproduced by $t_{1/2}(\rm{^{17}B})$, unlike the time spectrum
of 295(2)~keV, which is rather flat, suggesting a long lifetime of
the feeding $\beta$ decay. This flat spectrum is similar to the
242(1)-keV $\gamma$ line in Fig.~\ref{F:GMLIFE}(b), which is a BG
$\gamma$ emission line. With regard to the two points described
above, we concluded that the 217(2)- and 331(2)-keV $\gamma$ rays are
emitted subsequent to the $^{17}$B $\beta$ decay in the 0n channel,
while the 295(2)-keV $\gamma$ rays is emitted by long-lifetime
daughter isotopes produced in the $\beta$-decay chain from
$^{17}$B. The source of the 295(2)-keV $\gamma$ ray was not
identified due to its long lifetime, as compared with the present
value of the counting period $T_{\rm C}=$ 18~ms.

Finally, an $E_{\gamma}$ spectrum obtained from the
$\beta$-$\gamma$-$\gamma$ triple coincidence measurement was plotted
to determine whether the 217(2) and 331(2)-keV $\gamma$ rays can be
assigned to a direct transition to the GS or a cascade
transition. The second $\gamma$-ray coincidence was detected with the
set of thin plastic scintillators and NaI(Tl) detectors mentioned
above, wherein the signal from the plastic scintillator was used to
remove $\beta$-ray events. Figure~\ref{F:GAMMAgtd}(c) shows an
$E_{\gamma}$ spectrum obtained in coincidence with the second
$\gamma$ ray at $E_{\gamma} \leq {400}$~keV, measured with the
NaI(Tl) detectors. This energy value was selected to investigate the
cascade decay below the neutron threshold 
${S_{\rm n}}=0.729(18)$~MeV. Since no peaks were found at the energy
values $E_{\gamma}=$ 217(2) and 331(2)~keV in
Fig.~\ref{F:GAMMAgtd}(c), we concluded that these two $\gamma$ rays
are emitted from the  levels $E_{\rm x}=$ 217(2) and 331(2)~keV
during the transition to the GS in $^{17}$C.

\subsubsection{Other $\gamma$ lines}

\begin{figure}[tbh]
\begin{center}
 \includegraphics[width=.63\textwidth]{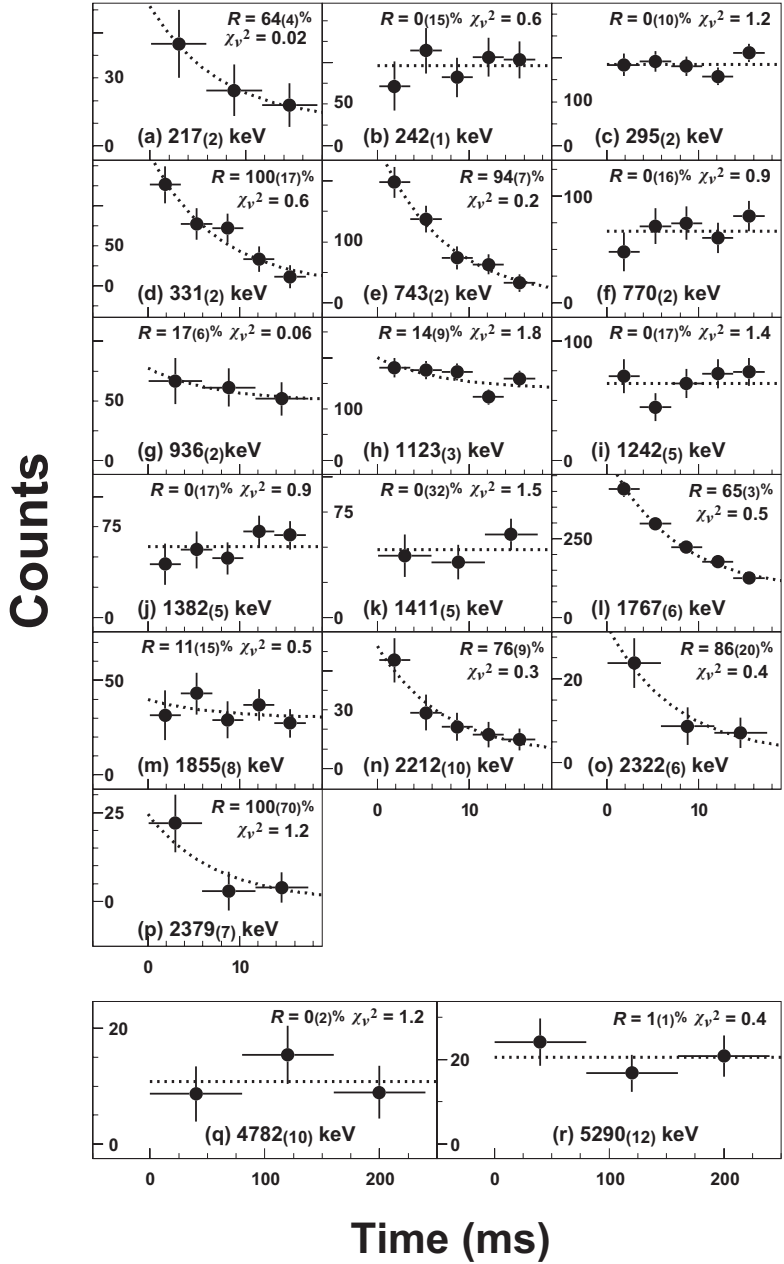}
 \caption{Time evolution of the photo-peak counts of the observed
  $\beta$-$\gamma$ lines measured with the Ge detector in the
  $^{17}$B $\beta$ decay. The vertical bar shows the statistical
  error, and the horizontal bar represents the time-slice window. The
  dotted line/curve shows the result of the $\chi^{2}$-fitting
  analysis using an exponential function with the
  known halflife of $^{17}$B, $t_{1/2}=$ 5.08(5)~ms, plus a constant
  is used. Values $R$ indicated in each panel shows the ratio of the
  exponential component to the total counts.
  The spectrum of (a) 217(2)-keV is obtained for an additional
  coincidence with a $\beta$ ray having $E_{\beta}>10$~MeV measured
  with the NaI(Tl) detector. Further, the spectra (q) and (r) are
  obtained with a measurement conducted separately by expanding
  $T_{\rm C}$ to 250~ms for investigating components with a long
  lifetime. 
\label{F:GMLIFE}}
\end{center}
\end{figure}

\begin{figure}[bthp]
\begin{center}
 \includegraphics[width=.80\textwidth]{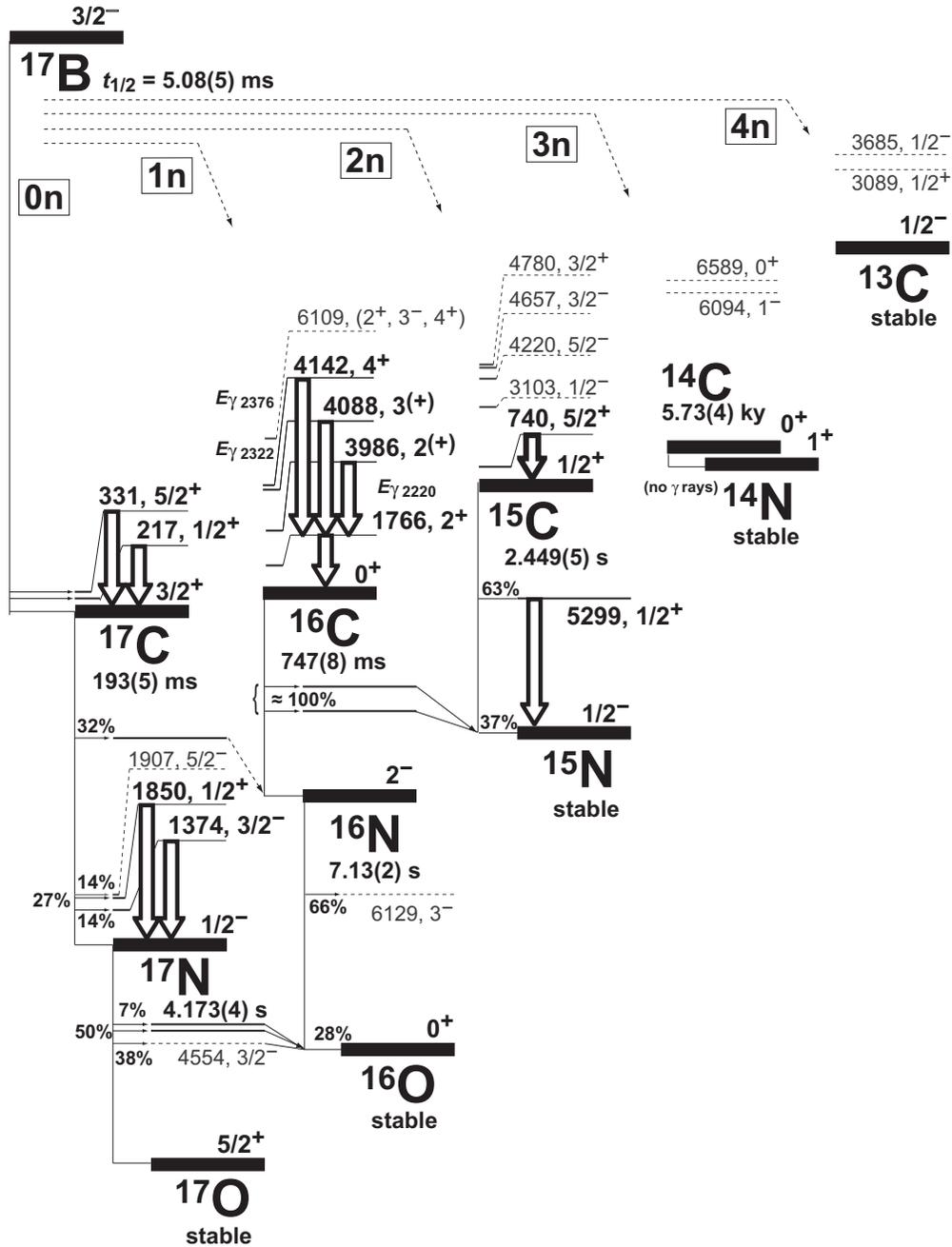}
 \caption{Observed $\gamma$ lines in the decay chain initiated by the
  $^{17}$B $\beta$ decay. The observed and identified $\gamma$ lines
  are shown by open arrows. 
\label{F:GM_OBS}}
\end{center}
\end{figure}

Other observed $\gamma$ lines, whose sources are not clear, are
plotted in Fig.~\ref{F:GMLIFE}. The time evolution of the $\gamma$
lines at $E_{\gamma}$= 743(3), 1767(6), 2212(10), 2322(6), and
2379(7)~keV, as shown in Figs.~\ref{F:GMLIFE}(e), (l), (n), (o), and
(p), respectively, are well reproduced by $t_{1/2}(\rm{^{17}B})$,
indicating that these $\gamma$ lines were directly fed by the
$^{17}$B $\beta$ decay, without the involvement of any daughter
isotope $\beta$ decays. As shown in Fig.~\ref{F:GM_OBS}, the
743(3)-keV $\gamma$ line was identified as the de-excitation $\gamma$
line of the $I^{\pi}=5/2^{+}$ state at $E_{\rm x}=743(2)$~keV
${\rightarrow}$ $1/2^{+}$ GS in $^{15}$C, while the $\gamma$ lines at
$E_{\gamma}=$ 1767(6), 2212(10), 2322(6), and 2379(7)~keV were
identified as the de-excitation $\gamma$ lines from the $E_{\rm x}=$
1766(10), 3986(7), 4088(7), and 4142(7)~keV states in $^{16}$C,
respectively. The $\gamma$ line at $E_{\gamma}=4782(10)$~keV observed
in Fig.~\ref{F:GAMMA}(c) agrees with the $\gamma$ line of
4780(100)~keV ${\rightarrow}$ GS for $^{15}$C. As shown in
Fig.~\ref{F:GMLIFE}(q), however, its time evolution is flat,
suggesting that the feeding $\beta$ decay is different from the
$^{17}$B $\beta$(-n) decay. Its source was not identified in this
study. In all $\beta$-decay channels associated with $\beta$-n
emission, the time evolution becomes flat for the given $T_{\rm C}$
window due to the long lifetimes of the daughter nuclei. The $\gamma$
lines at $E_{\gamma}=$ 1382(5) and 1855(8), shown in
Figs.~\ref{F:GMLIFE}(j) and (m), respectively, correspond to 
the de-excitation $\gamma$ lines at 1373.8(3) and 1849.5(3)~keV in
$^{17}$N, which are $\beta$-$\gamma$ decays of the daughter nucleus
$^{17}$C in the 0n channel of the $^{17}$B $\beta$ decay. Further, in
the 2n channel, the $\gamma$ line observed at 5290(12)~keV was
identified as the 5298.822(14)-keV $\gamma$ line in the $^{15}$C
$\beta$-$\gamma$ decay. Other peaks at $E_{\gamma}=$ 770(2), 936(2),
1123(3), and  1242(5)~keV were also investigated by studying their
time evolution, as shown in Figs.~\ref{F:GMLIFE}(f), (g), (h), and
(i), respectively;  no direct correlation of these peaks with the
$^{17}$B $\beta$ decay was observed. They were likely emitted from
the decay chain; however, the transitions that caused these peaks
could not be identified in this study.

\subsubsection{Transition strengths in the 0n channel}

Strengths $I_{\gamma}$ of the observed $\gamma$ lines per $^{17}$B
$\beta$ decay were determined using the following equation.  
%%------------------------------------------------------------------70
%\begin{widetext}
\begin{eqnarray}
I_{\gamma} 
= \frac{N_{\gamma}}
    {N({\rm ^{17}{\rm B}}) 
    \cdot {{\epsilon}_{\gamma}}{{\Omega}_{\gamma}}
    \cdot {{\epsilon}_{\beta}}{{\Omega}_{\beta}}
    \cdot R_{\beta}}
\label{Eq:Igamma}
\end{eqnarray}
%\end{widetext}
%%------------------------------------------------------------------70
Here, $N_{\gamma}$ is the photo peak count, $N$($^{17}$B) are
implanted $^{17}$B particles,
${{\epsilon}_{\gamma}}{{\Omega}_{\gamma}}$ and
${{\epsilon}_{\beta}}{{\Omega}_{\beta}}$ are, respectively, the
overall detection efficiencies of the Ge detector and the $\beta$-ray
telescopes, including their solid angles, which are evaluated based
on the simulation code of \textsc{GEANT}~\cite{GEANT}. $R_{\beta}$ is
the probability that a $\beta$ decay of a parent nucleus is directly
feeding a relevant $\beta$-$\gamma$-ray emission within the duration
$T_{\rm C}$. The $R_{\beta}$ values were obtained from a simulation
based on the following equation   
%%------------------------------------------------------------------70
%\begin{widetext}
\begin{eqnarray}
R_{\beta} =
\sum^{}_{i} \int^{{T_i}'}_{0}
{\rm d}t f(t;\lambda_1)
    F({T_i}' + t;\lambda_2)
    (1-F(T_{\rm C};\lambda_2)),\\
({T_i}' = {T_i} + T_{\rm R})\nonumber
\label{Eq:Rbeta}
\end{eqnarray}
%\end{widetext}
%%------------------------------------------------------------------70
where $f(t;\lambda) = \lambda \exp(-\lambda t)$ is the exponential
probability density distribution with a decay constant
$\lambda=1/\tau$, $F(t;\lambda)$ is its cumulative distribution
function, $\lambda_1$ is the inverse of the meanlife of the $^{17}$B
$\beta$ decay, $\lambda_2$ is that of the daughter's $\beta$-decay
constant that directly feeds a relevant $\gamma$-ray emission, and
${T_i}$ is the arrival time of $i$th $^{17}$B particle at which the
$T_{\rm{R}}+T_{\rm{C}}$ gate opens after the first implantation. The
probability distribution of ${T_i}$ is also given by
$f(t;\lambda_{\rm B})$ with the average $^{17}$B beam arrival rate
$\lambda_{\rm B}=17.3$ pps.

The transition strengths $I_{\gamma}$ of the $\gamma$ line, for which
decay schemes were identified, were determined using
Eq.~(\ref{Eq:Igamma}). The resulting $I_{\gamma}$ values, as well as
the corresponding values obtained after converting them to
$I_{\beta}$ and ${\log}{ft}$, are listed in
Table~\ref{T:0N_BRANCHES}.
In the derivation of the $I_{\beta}$ value associated with the
1767(6)-keV $\gamma$ line, the contributions from the cascade
transitions from $E_{\gamma}=$ 2212(10), 2322(6), and 2379(7)~keV
were subtracted. It should be noted that the $I_{\beta}$ values for
$E_{\gamma}=$ 2212(10), 2322(6), and 2379(7)~keV can include the sum
of other $\beta$-decay transitions in the 1n channel that are
followed by that particular $\gamma$ ray emission, because these
three $\gamma$ decays could be a part of cascade $\gamma$
decays from the higher $E_{\rm x}$ levels.

It is known that in the $^{17}$C $\beta$ decay, $\beta$-$\gamma$ rays
are emitted with $E_{\gamma}=$ 1373.8(3) and 1849.5(3)~keV with the
transition strengths $I_{\gamma}^{1374}(^{17}\rm{C})=24(8)$\% and and
$I_{\gamma}^{1850}(^{17}\rm{C})=22(5)$\%,
respectively~\cite{TIL93}. With the $I_{\gamma}$ values obtained
from the above analysis for the discussed $\gamma$ lines and the two
mentioned $I_{\gamma}(^{17}\rm{C})$ transition strengths, the total
0n-channel probability $P_{\rm{0n}}$ can be derived by using the
equation $P_{\rm 0n} = I_{\gamma}/I_{\gamma}(^{17}\rm{C})$. By taking
the weighted average of the resulting probabilities,
$P_{\rm{0n}}(\rm{1382(5)~keV}) = 25(9)$\% and 
$P_{\rm{0n}}(\rm{1855(8)~keV})=21(5)$\%,
we obtained $P_{\rm 0n} = 25(5)$\%. This probability agrees with the
reported $P_{\rm 0n}=$ 21(2)\%~\cite{DUF88}, within the assigned
error bounds.  

\subsection{1n-decay branches }
%%
%%==================================================================70

In the one-neutron (1n) decay channel, $^{17}$B decays into the
levels above the neutron threshold $S_{\rm n}=$ 0.729(18)~MeV. All of
these branches are connected with the GS of $^{16}$C. If a transition
is connected with an excited state in $^{16}$C, the neutron emission
is further followed by the de-excitation $\gamma$ ray emission. The
subsequent $\beta$ decay of $^{16}$C, whose halflife is
$t_{1/2}(^{16}{\rm C})=$ 747(8)~ms~\cite{AU97}, is similarly followed
by the neutron emission to $^{15}$N with $\sim$100\% branches in
all. This described decay cascade stops at the stable nucleus
$^{15}$N.

Thus, the 1n-decay channel has been studied through the measurement
of $\beta$-n~\cite{RAI96_17Bbn}, in which four excited states of
$^{17}$C, i.e., $E_{\rm x}=$ 1.18, 2.25, 2.64, and 3.82~MeV, were
identified using a TOF measurement, assuming the direct 1n-decay
transition to the $^{16}$C GS for all branches. In the present
measurement, the Ge and NaI(Tl) detectors were well tuned to detect
and thoroughly investigate de-excitation $\gamma$ rays in $^{16}$C
through $\beta$-n-$\gamma$ triple coincidence measurements. 

%%-------------------------------------------------------------------
%%
\subsubsection{TOF measurement of $\beta$-delayed neutrons}
%%
%%-------------------------------------------------------------------

\begin{figure}[bthp]
\begin{center}
 \includegraphics[width=.80\textwidth]{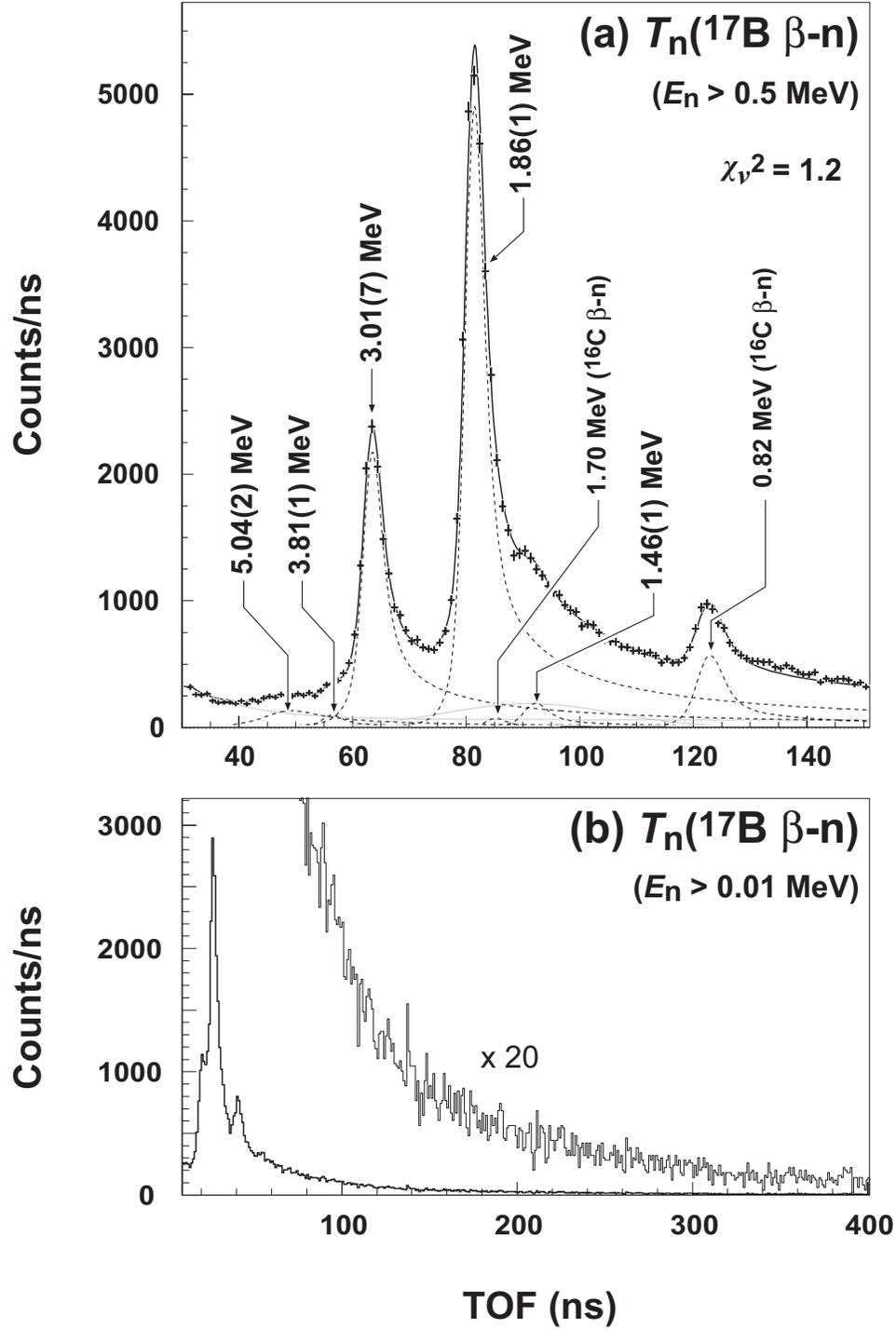}
 \caption{TOF spectra obtained for the $\beta$-delayed neutrons
  emitted in the $^{17}$B $\beta$ decay with (a) high-energy and (b)
  low-energy neutron detector arrays. The solid curve shown in panel
  (a) is the result of a least ${\chi}^{2}$-fitting  analysis The
  decomposed components are represented by dashed curves.
\label{F:17B-BETA-N}}
\end{center}
\end{figure}

Figure~\ref{F:17B-BETA-N} shows the obtained neutron TOF spectra for
the emitted neutrons from the $^{17}$B $\beta$ decay; they are
measured with (a) high-energy and (b) low-energy neutron detector
arrays, and the multiplicity of the neutron is limited to 
$M_{\rm n}=1$. As clearly seen in Fig.~\ref{F:17B-BETA-N}(a), three
peaks can be identified at 64, 81, and 122~ns, which correspond to
the neutron kinetic energies $E_{\rm n}=$ 3.01, 1.86, and 0.82~MeV,
respectively. Since it is known that in the $^{16}$C $\beta$ decay
($t_{1/2}=$ 747(8)~ms), $\beta$-n are emitted at $E_{\rm n}=$ 0.82(1)
and 1.70(1)~MeV~\cite{RAI96_17Bbn}, the observed 0.82-MeV peak
can be assigned to 0.82(1)~MeV. The other two peaks at $E_{\rm n}=$
3.01 and 1.86~MeV can be assigned to the $\beta$-n from the $^{17}$B
$\beta$ decay. These two major peaks were also observed at almost the
same energies $E_{\rm n}=$ 2.91(5) and 1.80(2) in
Ref.~\cite{RAI96_17Bbn}.

%%------------------------------------------------------------------70
%% FIG8 should be planced around here
%%------------------------------------------------------------------70
%  \begin{figure}[bthp]
%  \begin{center}
%   \includegraphics[width=.80\textwidth]{FIG8_UENO_NTOF.eps}
%   \caption{TOF spectra obtained for the $\beta$-delayed neutrons
%    emitted in the $^{17}$B $\beta$ decay with (a) high-energy and
%    (b) low-energy neutron detector arrays. The solid curve shown in
%    panel (a) is the result of a least ${\chi}^{2}$-fitting
%    analysis The decomposed components are represented by dashed
%    curves. 
%  \label{F:17B-BETA-N}}
%  \end{center}
%  \end{figure}
%%------------------------------------------------------------------70

The obtained neutron TOF spectrum shown in Fig.~\ref{F:17B-BETA-N}(a)
was analyzed using the response function (described later in the
text) of the neutron detector array. This response function was
determined so as to reproduce the well studied $\beta$-n from the
$^{17}$N $\beta$ decay~\cite{OWM76_17Nbn}. In the present analysis, the
following four factors were taken into account: (a) the level width
of the initial state, (b) uncertainty in the flight path of the
neutrons due to the counter misalignment and a finite beam-spot size,
(c) the long low-energy tail due to the scattering of emitted
neutrons from materials around the Pt stopper, and (d) uncertainty in
the time resolution, such as the intrinsic time resolution of the
$\beta$ and neutron detectors. In order to account for factor (a),
the response function essentially assumes a Lorentzian function
$H(t)$, and it includes the additional term $f(t)$ to account for
factor (c), where $f(t)$ is empirically determined. The response
function ${\cal F}(t)$ is then obtained as a  Gaussian convolution of
$H(t)+f(t)$ to account for factors (b) and (d). Thus, the function
${\cal F}(t)$ uses three parameters, the amplitude of peaks, the
neutron kinetic energy $E_{\rm n}$, and the level width $\Gamma$. The
detailed description of ${\cal F}(t)$ is given in
Ref.~\cite{MIYA03}.

Next, peak-decomposition was conducted through the least
$\chi^2$-fitting analysis with the obtained ${\cal F}(t)$
function. In this analysis, the following three types of BGs were
taken into account: (i) neutrons emitted in the multi-neutron
emission channel initiated by the $^{17}$B $\beta$ decay, which was
assumed to have a Gaussian shape with a wide width at the position
$\sim$100~ns, empirically determined from a TOF spectrum of neutrons,
measured with the neutron multiplicity $M \ge 2$; (ii) scattered
$\beta$ rays expressed by a spectral curve monotonically decreasing
with $t$; and (iii) a constant BG. The $\beta$-n emitted from the
$\beta$ decay of $^{16}$C, for $E_{\rm n}=$ 0.82(1) and 1.70(1)~MeV,
were also considered, and the reported relative intensities were
fixed in this case. The result of the fitting analysis is shown in
Fig.~\ref{F:17B-BETA-N}(a) using dashed curves. Other than the two
major peaks, which were determined to be at $E_{\rm n}=$ 3.01(1) and
1.86(1), the existence of minor peaks corresponding to neutron
kinetic energies $E_{\rm n}=$ 5.04(2), 3.81(1), and 1.46(1)~MeV was
suggested. Ref.~\cite{RAI96_17Bbn} has reported two minor transitions at
$E_{\rm n}=$ 1.43(2) and 0.42(1)~MeV. The first $E_{\rm n}$ value
listed agrees with the present $E_{\rm n}=$ 1.46(1) peak, with only a
small difference, although no peak corresponding to the second value
was visible in the spectrum. We also investigated neutrons at a lower
$E_{\rm n}$ with the low-energy neutron array. However, in the
obtained spectrum shown in Fig.~\ref{F:17B-BETA-N}(b), no further
peaks were identified at any  $E_{\rm n}$ down to 0.01~MeV.

%%-------------------------------------------------------------------
%%
\subsubsection{Neutron-decay branches to excited states in 
  $^{16}$C}
%%
%%-------------------------------------------------------------------

\begin{figure}[bthp]
\begin{center}
 \includegraphics[width=.80\textwidth]{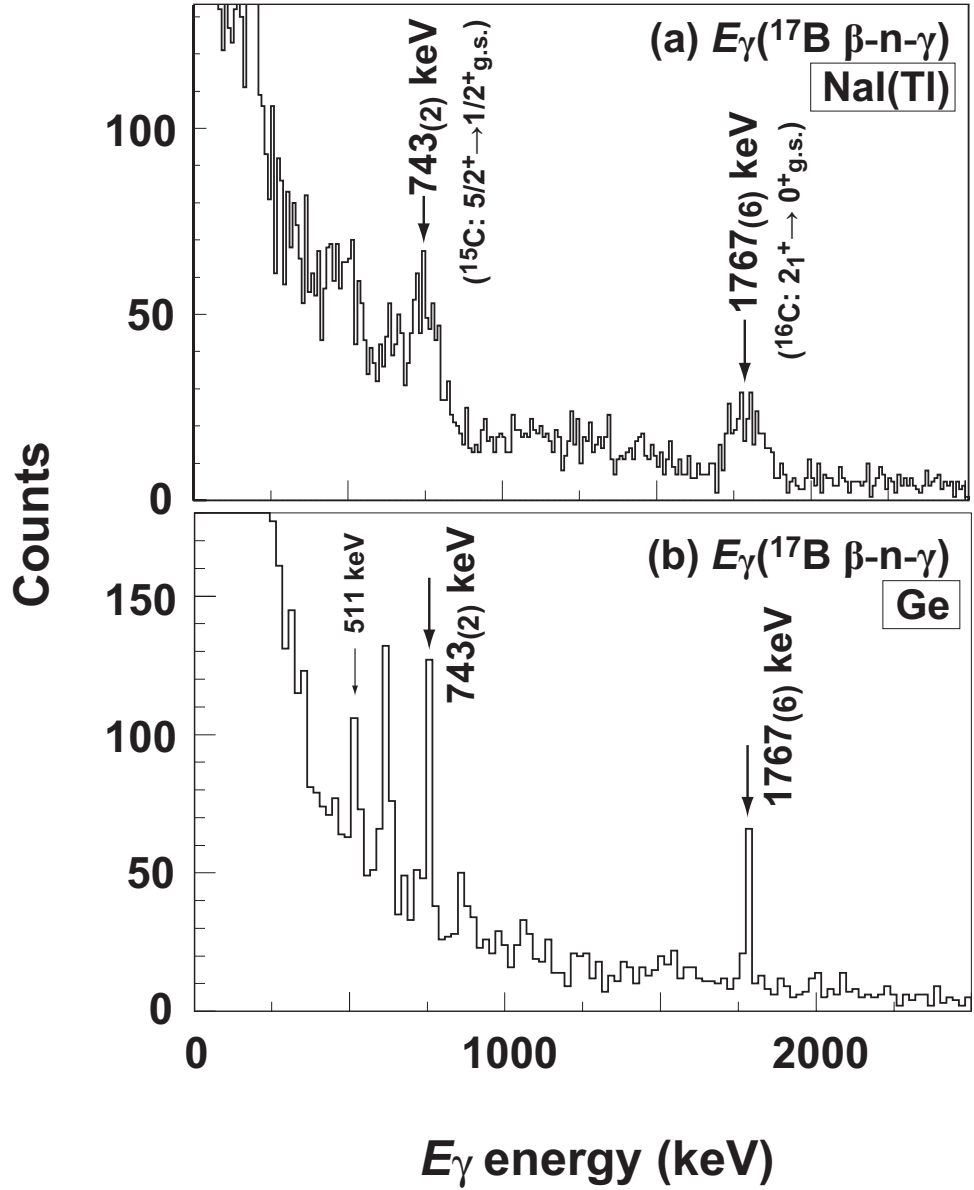}
 \caption{Obtained $\gamma$-ray energy spectra with (a) NaI(Tl)
  detectors and (b) the Ge   detector in coincidence with
  $\beta$-delayed neutrons with multiplicity $M_{\rm n}=$ 1. 
\label{F:GAMMA-N}}
\end{center}
\end{figure}

In order to construct the decay scheme, it is necessary to
investigate connected states in $^{16}$C  after neutron
emission. Here, we first studied a $\gamma$-ray energy spectrum
obtained in a $\beta$-n-$\gamma$ triple-coincidence measurement. As
shown in Figs.~\ref{F:GAMMA-N}(a) and (b), the de-excitation $\gamma$
rays from the first exited state in $^{16}$C at $E_{\rm x}=$
1766(10)~keV to the GS were clearly identified as a peak in the
$E_{\gamma}$ spectrum measured using both the NaI(Tl) and Ge
detectors. Another peak was also observed at $E_{\gamma}=$
743(2)~keV, which corresponds to the de-excitation $\gamma$ rays from
the $^{15}$C excited state at $E_{\rm x}=$ 740.0(15)~keV to its GS
following the two-neutron emission in the 2n channel of the $^{17}$B
$\beta$ decay. To confirm the initiating $\beta$ decay feeding the
two corresponding $E_{\gamma}$ emissions, the time evolution of their
counting rates were investigated as in the 0n-channel analysis. Both
the consequently obtained time spectra, shown in
Figs.~\ref{F:GMLIFE}(e) and (l), were well reproduced by
$t_{1/2}(\rm{^{17}B})$, which shows that these $\gamma$ emissions
were directly fed by the $^{17}$B $\beta$ decay. Further, two small
peaks are observed in Fig.~\ref{F:GAMMA}(b) at $E_{\gamma}=$ 2322(6)
and 2379(7), which agrees with the de-excitation $\gamma$ lines
2322(6) and 2379(7)~keV from the $3^{(+)}$ state at 
$E_{\rm x}=4088(7)$~keV and the $4^{+}$ state at 
$E_{\rm x}=4142(7)$~keV on $^{16}$C, respectively. The
time-evolution analysis described above was also performed for these
two $E_{\gamma}$ lines, and the results are shown in
Figs.~\ref{F:GMLIFE}(o) and (p), respectively. Although the
statistics are not sufficient, they are reproduced by
$t_{1/2}$($^{17}$B), suggesting that the $^{17}$B $\beta$ decay
directly feeds these two $\gamma$ emissions through $\beta$-n
emissions. Although peaks should appear at $E_{\gamma}=$ 2322(6) and
2379(7) in the gated $E_{\gamma}$ spectrum shown in
Fig.~\ref{F:GAMMA-N}, they are not observed due to the limited
statistics. 

%%------------------------------------------------------------------70
%% FIG9 should be planced around here
%%------------------------------------------------------------------70
%  \begin{figure}[bthp]
%  \begin{center}
%   \includegraphics[width=.80\textwidth]{FIG9_UENO_GMGTD1N.eps}
%   \caption{Obtained $\gamma$-ray energy spectra with (a) NaI(Tl)
%    detectors and (b) the Ge   detector in coincidence with
%    $\beta$-delayed neutrons with multiplicity $M_{\rm n}=$ 1. 
%  \label{F:GAMMA-N}}
%  \end{center}
%  \end{figure}
%%------------------------------------------------------------------70

\begin{figure}[bthp]
\begin{center}
 \includegraphics[width=.80\textwidth]{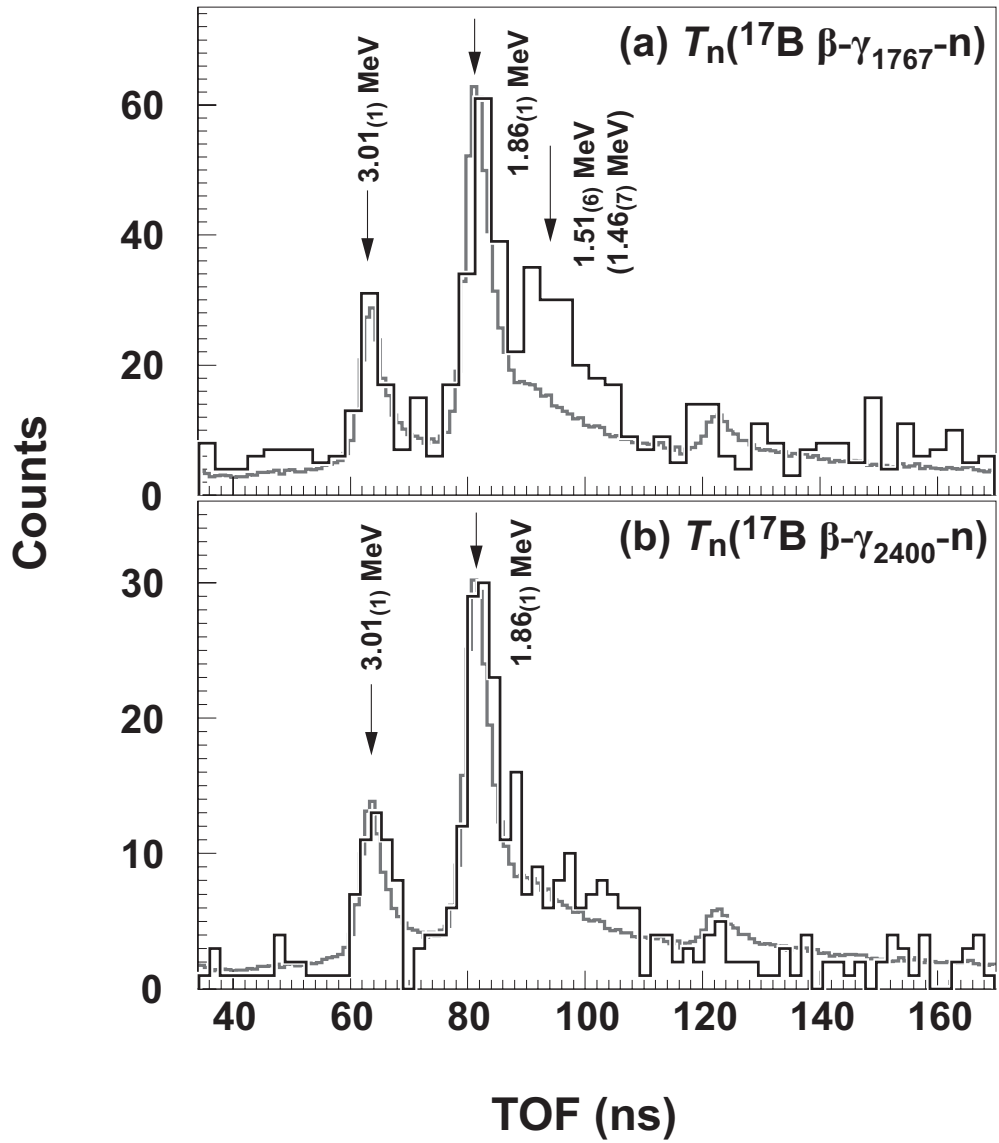}
\caption{TOF spectrum obtained with the neutron detector array in
  coincidence with a 1767(6)-keV $\gamma$ peak observed with NaI(Tl)
  detectors.
\label{F:MAGRgwNAI}}
\end{center}
\end{figure}

In the next step, $\beta$-n components connected to the 
$E_{\rm x}=1766(10)$~keV state in $^{17}$C were
investigated. Figure~\ref{F:MAGRgwNAI} shows a neutron TOF spectrum
measured under the $\beta$-n-$\gamma$ coincidence condition, and in
the measurement $E_{\gamma}$ energies were set as $E_{\gamma} \simeq$
(a) 1767~keV and (b) 2400~keV. To enable clear comparison, the
normalized TOF spectrum without the $E_{\gamma}$ window (i.e., the
spectrum shown in Fig.~\ref{F:17B-BETA-N}(a)) is also drawn using
gray lines. In both the spectra, two major peaks at $E_{\rm n}=$
3.01(1) and 1.86(1)~MeV still appear. However, this does not imply
that the two peaks are connected to the 1767(6)-keV $\gamma$
line. The peaks are caused by accidental coincidences owing to the
following two reasons. First, The TOF spectrum obtained for
$E_{\gamma}=2400$~keV, at which no Compton scattering from the
$E_{\gamma}=$ 1767(6)~keV $\gamma$ rays can interfere and no other
photo peaks are observed, is analogous to the spectrum in the ungated
case, as shown in Fig.~\ref{F:MAGRgwNAI}(b). In addition, the
obtained counting rates of the two major peaks are $\sim$20 times
smaller than those expected on the basis of the spectrum shown in
Fig.~\ref{F:17B-BETA-N}(a) and the given NaI(Tl) efficiency. If we
closely examine Fig.~\ref{F:MAGRgwNAI}(a), however, a peak can be
observed at $\sim$95~ns, which is emphasized by the $E_{\gamma}=$
1767(6)~keV gate. This peak energy, $E_{\rm n}=1.51(6)$~MeV, agrees
with the  $E_{\rm n} =1.46(1)$~MeV obtained in the described peak
decomposition analysis of the ungated $E_{\rm n}$ spectrum. In
addition, the transition strength $I_{\beta}=1.7(2)$\%, evaluated
using the 1.51-MeV peak count based on Eq.~(\ref{EQ:Ibeta_1n}), which
accounts for the NaI(Tl) efficiency, agrees with the value
$I_{\beta}=1.5(2)$\% for the 1.46-MeV peak  determined as described
in Sec.~\ref{Sec:Ibeta}. It is then natural to conclude that both the
peaks are identical. No peaks are observed at $E_{\rm n} =$ 5.04(2)
and 3.81(1)~MeV in Fig.~\ref{F:MAGRgwNAI}(a), although peaks with
statistics comparable to the $E_{\rm n} = 1.46(1)$ peak should be
observed if they are also connected to the $E_{\rm x}=1766(10)$~keV
state. 

%%------------------------------------------------------------------70
%% FIG10 should be planced around here
%%------------------------------------------------------------------70
%  \begin{figure}[bthp]
%  \begin{center}
%   \includegraphics[width=.80\textwidth]{FIG10_UENO_NTOFGTD.eps}
%  \caption{TOF spectrum obtained with the neutron detector array in
%    coincidence with a 1767(6)-keV $\gamma$ peak observed with
%    NaI(Tl) detectors.
%  \label{F:MAGRgwNAI}}
%  \end{center}
%  \end{figure}
%%------------------------------------------------------------------70

%%-------------------------------------------------------------------
%%
\subsubsection{Neutron-decay branches to the $^{16}$C ground state}
%%
%%-------------------------------------------------------------------

\begin{figure}[bthp]
\begin{center}
 \includegraphics[width=.80\textwidth]{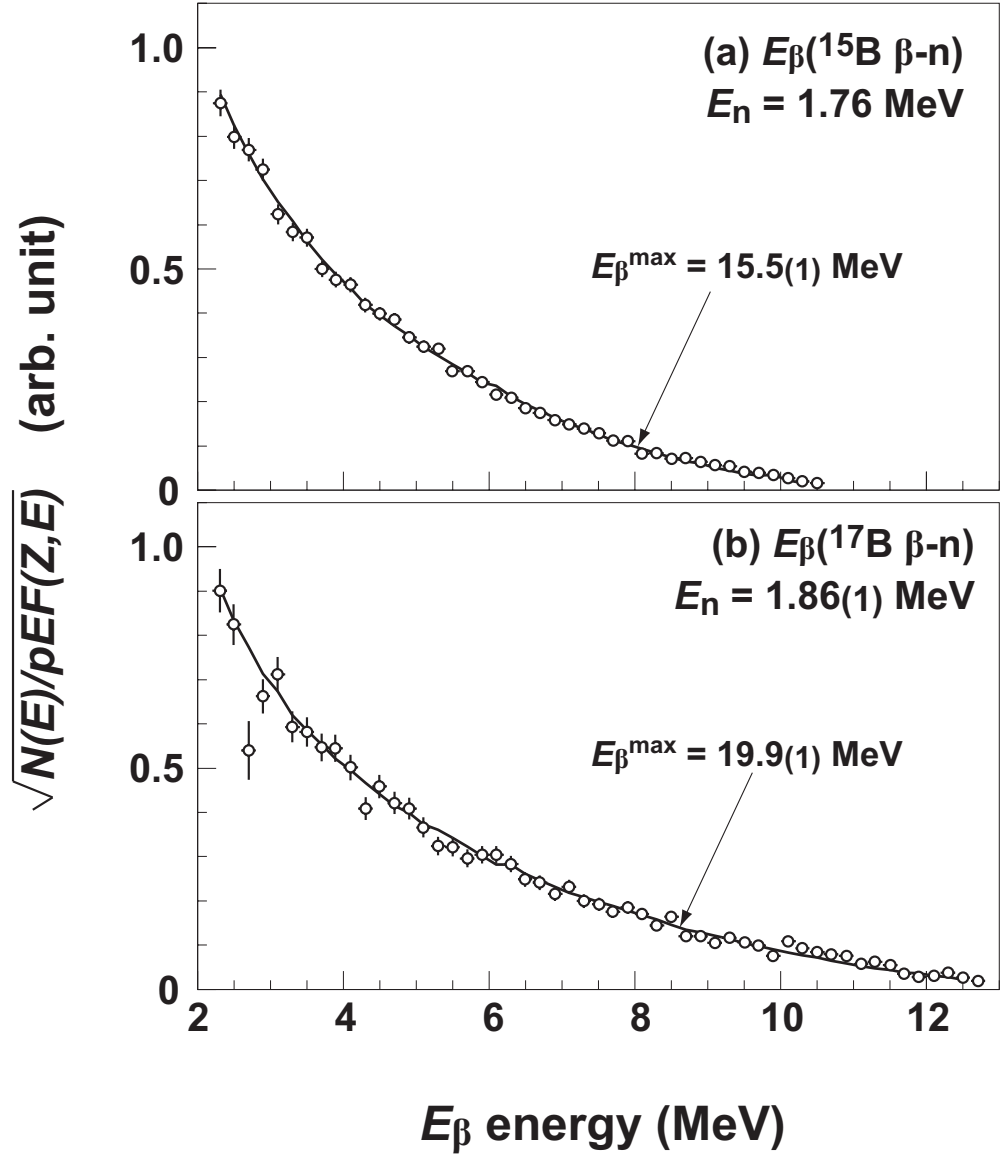}
\caption{$\beta$-Ray energy spectra (i.e., Kurie plot) measured with
  NaI(Tl) detectors in coincidence with (a) the 1.86(1)-MeV
  $\beta$-delayed neutrons in the $^{17}$B $\beta$ decay and (b)
  1.76-MeV $\beta$-delayed neutrons in the $^{15}$B $\beta$
  decay. Solid curves show the result of \textsc{GEANT} simulations,
  in which the end-point energies $E_{\beta}^{\rm max}=$  19.7(1)~MeV
  and 17.9(2)~MeV were deduced, respectively. For comparison with the
  simulation results, $\beta$-ray yields are normalized to unity at
  $E_{\beta}=$ 2~MeV. 
\label{F:BETA}}
\end{center}
\end{figure}

Because it was shown that only the 1.46(1)-MeV peak was connected to
the 1767(6)-keV $\gamma$ line, the two major peaks at $E_{\rm n} =$
3.01(1) and 1.86(1)~MeV, as well as the minor peaks at $E_{\rm n} =$
5.0 and 3.8~MeV, were assigned to the direct transition to the
$^{16}$C GS. As described in the following text, this assignment was
verified with the $E_{\rm n} = 1.86(1)$~MeV peak having the highest
statistics, through the measurement of the end-point energy
$E_{\beta}^{\rm max}$ of the $\beta$ decay associated with this
$E_{\rm n}$ peak. Figure~\ref{F:BETA}(b) shows a $\beta$-ray energy
spectrum of the $^{17}$B $\beta$ decay obtained
with the combination of the NaI(Tl) and plastic scintillators in the
$\beta$-n measurement by gating at $E_{\rm n}=$ 1.86(1)~MeV. 
The spectrum is expressed in the form of the Kurie plot
$\sqrt{N(E)/pEF(Z,E)}$, where $E$ and $p$ are of the total energy and
momentum of the $\beta$ particle, $N(E)$ is the observed $\beta$-ray
yield at $E$, $F(Z,E)$ is the Fermi function, and $Z$ is the atomic
number of the parent nucleus. The obtained $\beta$-ray energy
spectrum was analyzed using a \textsc{GEANT}~\cite{GEANT} simulation,
in which an arrowed $\beta$-decay transition was assumed by taking
into account its ${\log}{ft}$ value 4.8(1), determined in the present
work, as explained later. The obtained end-point energy
$E_{\beta}^{\rm max}=$ 19.7(1)~MeV agrees well with the 19.9(1)~MeV
value for the transition to the $^{16}$C GS calculated using the
$Q_{\beta}$ of $^{17}$B and the neutron energy $E_{\rm n}=$
1.86(1)~MeV, while it differs from $E_{\beta}^{\rm max}=18.1(1)$~MeV
when the $\beta$-n emission is connected to the 1767(6)~keV $\gamma$
emission line. The accuracy of the analysis was evaluated using the
$^{15}$B $\beta$ decay measured in the present work. It is known that
in the $^{15}$B $\beta$ decay, the transition to $E_{\rm
  x}=3.10$~MeV, which has the highest transition strength in the
$\beta$ decay, is followed by the  $E_{\rm n}=1.76$~MeV $\beta$-n 
emission~\cite{HAR99_15B,MIYA03}. Thus, a Kurie plot was also
obtained for this transition using the same experimental
apparatus. As shown in Fig.~\ref{F:BETA}(a), the end-point energy
$E_{\beta}^{\rm max}=15.5(1)$~MeV obtained through the same procedure
again agrees with the calculated value  
$E_{\beta}^{\rm max}=16.0(1)$~MeV, with only a slight difference in
values. Due to the limited statistics, this analysis cannot be
carried out for the remaining minor peaks at $E_{\rm n}=$ 5.0(3) and
3.8(1)~MeV.
%%------------------------------------------------------------------70
%% FIG11 should be planced around here
%%------------------------------------------------------------------70
%  \begin{figure}[bthp]
%  \begin{center}
%   \includegraphics[width=.80\textwidth]{FIG11_UENO_KURIEPLT.eps}
%   \caption{$\beta$-Ray energy spectra (i.e., Kurie plot) measured
%     with NaI(Tl) detectors in coincidence with (a) the 1.86(1)-MeV
%     $\beta$-delayed neutrons in the $^{17}$B $\beta$ decay and (b)
%     1.76-MeV $\beta$-delayed neutrons in the $^{15}$B $\beta$
%     decay. Solid curves show the result of \textsc{GEANT}
%     simulations, in which the end-point energies $E_{\beta}^{\rm
%       max}=$  19.7(1)~MeV and 17.9(2)~MeV were deduced,
%     respectively. For comparison with the simulation results,
%     $\beta$-ray yields are normalized to unity at $E_{\beta}=$
%     2~MeV. 
%  \label{F:BETA}}
%  \end{center}
%  \end{figure}
%%------------------------------------------------------------------70

\subsubsection{Transition strengths in the 1n channel}

As mentioned above, the $^{17}$B $\beta$ decay in the 1n channel is
100\%, followed by the successive $\beta$-n emission in the $\beta$
decay of the daughter nucleus of $^{16}$C. Because in the $^{16}$C
$\beta$ decay, the strength of the transition feeding the $\beta$-n
emission at $E_{\rm n}=0.82(1)$ is known to be $I_{\beta}=$
84.4(17)\%, the total 1n decay strength  $P_{\rm{1n}}$ of the
$^{17}$B $\beta$ decay can be determined by 
%%------------------------------------------------------------------70
%\begin{widetext}
\begin{eqnarray}
P_{\rm 1n}
= \frac{N_{\rm n}}
    {N({\rm ^{17}{\rm B}})\cdot I_{\beta}
    \cdot {{\epsilon}_{\rm n}}{{\Omega}_{\rm n}}
    \cdot {{\epsilon}_{\beta}}{{\Omega}_{\beta}}
    \cdot R_{\beta}},
\label{Eq:In}
\end{eqnarray}
%\end{widetext}
%%------------------------------------------------------------------70
where ${N_{\rm n}}$ is the count of the 0.82(1)-MeV peak obtained in
the described ${\chi}^2$-fitting analysis and ${\epsilon}_{\rm n}$
and ${{\Omega}_{\rm n}}$ are the efficiency and the solid angle of
the neutron counter array for the 0.82(1)-MeV neutron. The values of
${\epsilon}_{\rm n}$ were evaluated based on Ref.~\cite{CEC79}, using
experimental ${\epsilon}_{\rm n}$ data obtained from the measurements
separately conducted with $^{17}$N and $^{15}$B beams, as described
in Sec.~\ref{S:NeutEff}. Here, the ${\epsilon}_{\rm n}$ of the
$\beta$-n at $E_{\rm n}=$ 1.17~MeV in the $\beta$ decay of $^{17}$N,
which is close to 0.82(1)~MeV, was determined with an accuracy of
$\Delta{\epsilon}_{\rm n}=9.4$\%. The same $\Delta{\epsilon}_{\rm n}$
value was therefore adopted in the evaluation of ${\epsilon}_{\rm n}$
for the 0.82(1)-MeV neutron as a systematic error. We thus obtained
$P_{\rm{1n}}=67(7)$\%, which agrees with the reported value
$P_{\rm{1n}}=63(1)$\%~\cite{DUF88}, within the assigned error
bounds.

However, for the other $E_{\rm n}$ energies of observed $\beta$-n,
the accuracies of the obtained ${\epsilon}_{\rm n}$ reference data
were not sufficiently high. Therefore, the 1n transition strengths
were determined by solving for the  normalization factor $\alpha$ in
the following equation  
%%------------------------------------------------------------------70
\begin{eqnarray}
P_{\rm 1n} =
\alpha \left( \sum^{}_{i} I'_{\beta}(i) 
  - I'_{\beta}(1.46~{\rm~MeV}) \right)  + 9.4\%,
\label{EQ:Ibeta_1n}
\end{eqnarray}
%%------------------------------------------------------------------70
\label{Sec:Ibeta}
where $I'_{\beta}(i)$ is the relative transition strength obtained
with a relative ${\epsilon}_{\rm n}$ value for each transition $i$
and 9.4\% is the sum of the strengths of the 1n transitions that are
finally followed by $E_{\gamma}=$ 1767(6)~keV $\gamma$ ray emissions,
which were not included in the left-hand side in
Eq.~(\ref{EQ:Ibeta_1n}), with the exception of
$I'_{\beta}(1.46~{\rm~MeV})$. The resultant $I_{\beta}$ values
calculated using ${\alpha}{I'_{\beta}}$, as well as the values
obtained by converting them to ${\log}{ft}$ and $B$(GT), are
summarized in Table~\ref{T:1N_BRANCHES}.

\subsection{2n-decay and the higher-multiplicity channels} 
%%
%%==================================================================70

The $^{17}$B $\beta$ decay in the 2n-decay channel feeds states in
$^{15}$C through the $\beta$-2n emission. All 2n-decay branches are
then followed by the $\beta$ decay of $^{15}$C ($t_{1/2}=$
2.449(5)~s~\cite{ALB79_t15C}), where the $E_{\gamma}=$ 5290(12)~keV
$\gamma$ ray is expected be emitted from the $I^{\pi}=$ 1/2$^+$
excited state at $E_{\rm x}=$ 5298.822(14)~keV in $^{15}$N, with the
branching ratio $I_{\beta}^{5299}({\rm ^{15}C}) =
63.2(8)$\%~\cite{AJ91}. The corresponding peak was observed at
$E_{\gamma} =$ 5290(12)~keV in the obtained Ge spectrum, as shown in
Fig.~\ref{F:GAMMA}(c), for which the deduced transition strength
$I_{\gamma}({\rm ^{17}B}) = 7.5(11)$\% is listed in
Table~\ref{T:0N_BRANCHES}. The total 2n-decay probability, 
$P_{\rm 2n}$, can then be determined  as 
$P_{\rm 2n}=$ 
$I_{\gamma}^{5290(12)}({\rm{^{17}B}})/I_{\beta}^{5299}({\rm{^{15}C}})$
$=$ 12(2)\%. It should be noted that this $P_{\rm 2n}$ value agrees
well with the value $P_{\rm 2n}=$ 11(7)\% reported in
Ref.~\cite{DUF88}, within the assigned error bounds.

In the 2n channel, the $\gamma$ ray observed at $E_{\gamma}=$
743(2)~keV in Fig.~\ref{F:GAMMA}(a) can be assigned to the
de-excitation $\gamma$ ray from the  $I^{\pi}=$ 5/2$^{+}$ first
excited state at $E_{\rm x}=$ 740.0(15)~keV~\cite{AJ91} to the
$I^{\pi}=$ 1/2$^{+}$ GS of $^{15}$C. This assignment was confirmed by
observing that the time evolution of the photo-peak counts, shown in
Fig.~\ref{F:GMLIFE}(e), can be well reproduced by
$t_{1/2}(\rm{^{17}B})$. From the measured peak count, the sum of the
strengths of the $^{17}$B $\beta$-decay transitions connected to the
$E_{\rm x}=740.0(15)$~keV state of $^{15}$C was determined to be
$I_{\beta}=2.6(2)$\%, as listed in Table~\ref{T:0N_BRANCHES}. Since
no $\gamma$ lines of $^{15}$C, other than the one at 743(2)~keV, were
identified in the $E_{\gamma}$ spectrum, we concluded that the
740.0(15)-keV state is directly connected to the two-neutron emission
and that the GS of $^{15}$C is directly fed by the strength
$P_{\rm{2n}}-I_{\beta}= 9.4(3)$\%. In order to construct the decay
scheme of the 2n-decay channel, it is necessary to identify the sum
energy of the two emitted neutrons. However, the obtained statistics
provided no clear evidence of the 2n decay.

%% 3n 4n 
With respect to the 3n(4n)-decay branch, the $\beta$-n emission from
$^{17}$B feeds states in $^{14}$C($^{13}$C), for which the total
transition probability is reported to be  $P_{\rm 3n}=$ 3.5(7)\%
($P_{\rm 4n}=$ 0.4(3)\%)~\cite{DUF88}. Because the halflife of
$^{14}$C is $t_{1/2}=$ 5730 y and $^{13}$C is stable, the
$\beta$-decay cascade from $^{17}$B is almost stopped at the GSs. In
the measurement, clear peaks were not found in the sum-energy spectra
of neutrons with multiplicity $M_{\rm n}=$ 3 or 4 with the present
statistics. Furthre, no $\gamma$-ray peaks were found at the
$E_{\gamma}$ energies associated with the de-excitation $\gamma$ rays
in $^{13}$C and $^{14}$C. Thus, both $P_{\rm 3n}$ and $P_{\rm 2n}$
were not determined in the present work. 

%%==================================================================70
%%
\subsection{Spin-parity assignment}
%%
%%==================================================================70

\begin{figure}[bthp]
\begin{center}
 \includegraphics[width=.80\textwidth]{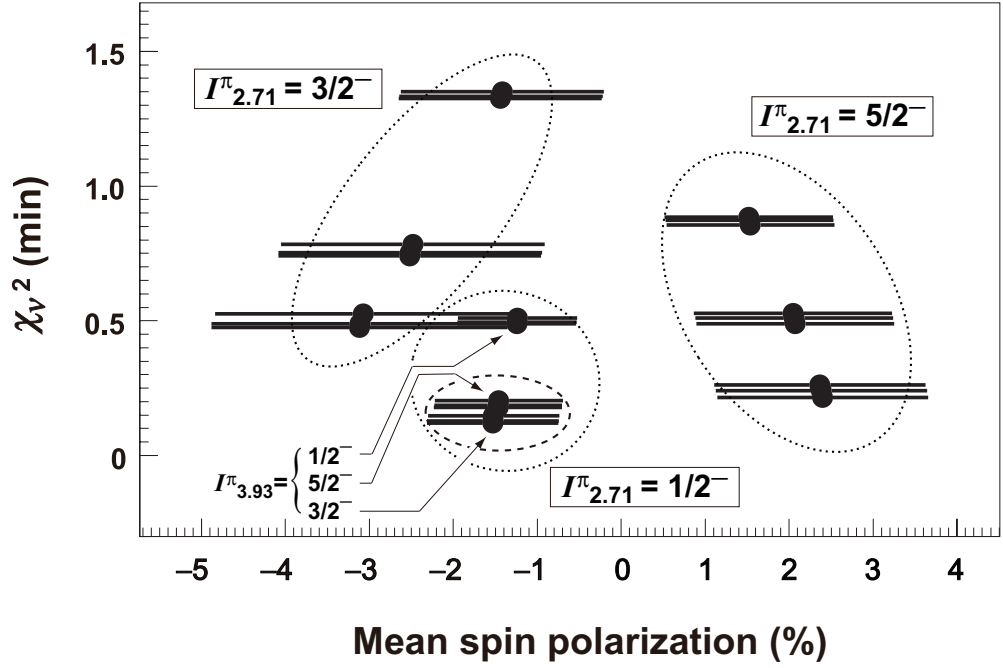}
\caption{Values of $\chi_{\nu}^{2}$ plotted as a function of the mean
  spin polarization calculated for all possible sets of $A_{\beta}$
  values in the $^{17}$B $\beta$-decay transition to the observed
  levels in $^{17}$C at $E_{\rm x}=$ 2.71(2), 3.93(2), and
  4.05(2)~MeV. Classification according to the $I^{\pi}$ for $E_{\rm
    n}=$ 2.71(2)~MeV is indicated by dotted ellipses. The dashed
  ellipse shows the most probable sets of the $A_{\beta}$ allocation. 
\label{F:APFIT}}
\end{center}
\end{figure}

After obtaining the peak counts of the $\beta$-n TOF spectra in the
peak-decomposition analysis, $A_{\beta}P$ values of the transition to
$E_{\rm x}=$ 2.71(2), 3.93(2), and 4.05(2)~MeV, for which two
unassigned peaks were removed, were determined based on
Eq.~(\ref{EQ:AP}), and the values are listed in
Table~\ref{T:1N_BRANCHES}. In order to determine $I^{\pi}$ values,
the least $P$-variance method described in Sec.~\ref{S:SP_ASSIGN}
was applied to the obtained $A_{\beta}P$ values. For all possible
$3^{3}=27$ combinations of $I^{\pi}$, $\chi_{\nu}^{2}$, given by
Eq.~(\ref{EQ:chisq}), were calculated together with the corresponding
mean spin polarization $\bar{P}$, defined in
Eq.~(\ref{EQ:Pmean}). The resulting $\chi_{\nu}^{2}$ values are
plotted as a function of $\bar{P}$ in Fig.~\ref{F:APFIT}. The
obtained $\chi_{\nu}^{2}$ values can be divided into three groups, as
indicated by the dotted ellipses in Fig.~\ref{F:APFIT}: 
$\bar{P} \simeq$  $-2.5$, $-1.5$, and $+2.4$\%. This classification
is based on the $I^{\pi}_{\rm f}$ value of the 2.71(2)-MeV state
owing to its highest statistics.

We first note that a negative spin polarization is expected for the
$^{17}$B beam produced in the present experiment, based on the
systematics observed with similar beam energies, targets, and
projectile-ejectile
combinations~\cite{ASAHI,OKUNO,OGW02_17C,HU17B,30AL32AL,17BQ,18N,%%%%
1415BQ,1415B_MYU}, for which the reaction proceeds through a far-side
trajectory owing to the nuclear attractive force dominating the
Coulomb repulsive force. Actually, the negative spin polarization of
$^{15}$B, $\bar{P}=-2.81(42)$\%, was observed with the same beam
energies, targets, and projectile~\cite{MIYA03}. Thus, the
possibility of assigning $5/2^{-}$ to the 2.71(2)-MeV state can be
eliminated owing to its positive polarization $\bar{P} \simeq
+2.4$. Among the remaining two $I^{\pi}$ values, $1/2^{-}$ and
$3/2^{-}$, the former can be assigned using the obtained minimum
$\chi_{\nu}^{2}$ analysis, although the possibility that $3/2^{-}$
can be assigned is not possible to completely ignore.

After we assigned $1/2^{-}$ to the 2.71(2)-MeV state, the $I^{\pi}$
combinations could be further divided into two groups: (i) $1/2^{-}$
and (ii) $3/2^{-}$ or $5/2^{-}$ assignments for the 3.93(2)-MeV
state. The $\chi_{\nu}^{2}$ values of the first group are in the 
range of $0.49 \sim 0.51$, whereas the second group has a smaller
$\chi_{\nu}^{2} = 0.12 \sim 0.26$. We thus assigned $(3,\ 5)/2^{-}$
to the 3.93(2)-MeV state. However, no further $I^{\pi}$ selection was
performed, because the differences among the six combinations in the
second group were extremely small. The $I^{\pi}$ assignment completed
in this analysis is listed in Table~\ref{T:1N_BRANCHES}, together
with the asymmetry parameters calculated from the expected
polarization $\bar{P}=$ 1.5(8) and given by the minimum
$\chi_{\nu}^{2}$ set. 
　
%%------------------------------------------------------------------70
%% FIG12 should be planced around here
%%------------------------------------------------------------------70
%  \begin{figure}[bthp]
%  \begin{center}
%   \includegraphics[width=.80\textwidth]{FIG12_UENO_AP.eps}
%   \caption{Values of $\chi_{\nu}^{2}$ plotted as a function of the 
%    mean spin polarization calculated for all possible sets of
%    $A_{\beta}$ values in the $^{17}$B $\beta$-decay transition to
%    the observed levels in $^{17}$C at $E_{\rm x}=$ 2.71(2),
%    3.93(2), and 4.05(2)~MeV. Classification according to the
%    $I^{\pi}$ for $E_{\rm n}=$ 2.71(2)~MeV is indicated by dotted
%    ellipses. The dashed ellipse shows the most probable sets of the
%    $A_{\beta}$ allocation.  
%  \label{F:APFIT}}
%  \end{center}
%  \end{figure}
%%------------------------------------------------------------------70

%%%%%%%%%%%%%%%%%%%%%%%%%%%%%%%%%%%%%%%%%%%%%%%%%%%%%%%%%%%%%%%%%%%%70
%%
%%
\section{Discussion}
%%
%%
%%%%%%%%%%%%%%%%%%%%%%%%%%%%%%%%%%%%%%%%%%%%%%%%%%%%%%%%%%%%%%%%%%%%70

\subsection{Further investigation for $I^{\pi}$ assignments}
The obtained level structure of $^{17}$C was further investigated for
comparison with the $\beta$ decays of $^{13}$B and $^{15}$B. The
several negative-parity states in $^{13}$C ($^{15}$C), connected with
the $\beta$ decay of $^{13}$B ($^{15}$B), are illustrated in
Fig.~\ref{F:SHELLMODEL}, where relative $B{\rm (GT)}$ strengths
calculated from the reported ${\log}{ft}$ values~\cite{AJ91} are
indicated by the line thickness, assuming that Fermi transition
strengths are negligibly small owing to the large difference between
their ${E_{\beta}}^{\rm max}$ energies and $Q_{\beta}$ values. In the
$\beta$ decay of $^{13}$B, the transition strength is concentrated at
the $\beta$-decay branch leading to the $1/2_{1}^{-}$ GS of $^{13}$C,
whose branching  ratio is 92.1\%. Including the second strongest
branch with the strength 7.6\%, leading to the $3/2_{1}^{-}$ excited
state at $E_{\rm x}=$ 3684.507(19)~keV, the sum of the $^{13}$B
$\beta$-decay transitions exhausts $\sim$100\% of the total transition
strength. Because the main GS configuration of $^{13}$B can be
described by 
$|^{13}{\rm B_{GS}}{\rangle}^{3/2^-}
= |(\pi p_{3/2})_{-1} \otimes
(\nu p_{3/2})_{4}^{0^+}(\nu p_{1/2})_{2}^{0^{+}}{\rangle}^{3/2^{-}}$, 
given the well-established neutron magic number $N=8$,
% the concentration of the transition strength for the transition to
the transition to the $1/2_{1}^{-}$  GS
can be described by the single-particle GT transition 
$\nu p_{1/2}\rightarrow\pi p_{3/2}$, 
under the reasonable assumption of a configuration
$|^{13}{\rm C_{GS}}\rangle ^{1/2^{-}}
= | (\pi p_{3/2})_{4}^{0^{+}} \otimes
(\nu p_{3/2})_{4}^{0^{+}}(\nu p_{1/2})_{1}{\rangle}^{1/2^{-}}$ to the
$1/2_{1}^{-}$ state.
%%-------new sentenses-------------->
Then, the concentration of the transition strength to the
$1/2_{1}^{-}$ state can be understood taking 
this pure single ${\nu}{p_{1/2}}$-particle state to the $1/2_{1}^{-}$
state, because the $\nu p_{1/2}\rightarrow\pi p_{3/2}$ GT transition
can only populate this configuration from the $^{13}$B GS.
%%<--------------new sentenses-------
A similar observation applies to the $^{15}$B $\beta$ decay, in which
the lowest negative-parity state at $E_{\rm x}=3.103$~MeV~\cite{AJ91}
in $^{15}$C, which has been assigned to
$I^{\pi}={1/2^{-}}$~\cite{HAR99_15B,MIYA03}, is fed through the
strongest GT transition. Assuming a weak coupling of the $^{13}$B GS
with the two valence  neutrons in the $sd$ orbit, the $^{15}$B GS can
be written as 
$ |^{15}{\rm B_{GS}}{\rangle}^{3/2^{-}}
= |^{13}{\rm B_{GS}}{\rangle}^{3/2^{-}} \otimes
|({\nu}sd)_{2}{\rangle}^{0^{+}}$.
Then, the observed analogous characteristics of the GT transitions in
the $\beta$ decays of $^{13}$B and $^{15}$B can be understood,
provided a major contribution of the GT transition from $^{15}$B to
the lowest ${1/2^{-}}$ state in $^{15}$C is largely governed by the 
${\nu}{p_{1/2}}{\rightarrow}{\pi}{p_{3/2}}$ transition
in the $^{13}$B GS,
where the two neutrons are kept in the $sd$ orbit to
couple to form 
$I^{\pi}=0^{+}$. For detailed discussions, such as those on the
inversion of the  $3/2_{1}^{-}$ and $5/2_{1}^{-}$ states in $^{15}$C,
effects of the $p$-$sd$ cross-shell interactions need to be taken
into account. The same rule can be applied to the $^{17}$B $\beta$
decay, because the main configuration of the $^{17}$B GS should be
given by  
 $|^{17}{\rm B_{GS}}{\rangle}^{3/2^{-}}
= |^{13}{\rm B_{GS}}{\rangle}^{3/2^{-}} \otimes
  |({\nu}sd)_{4}{\rangle}^{0^{+}}$.
Then, the strongest GT transition can be characterized by
${\nu}{p_{1/2}}{\rightarrow}{\pi}{p_{3/2}}$ 
in the $^{13}$B GS, 
and the final state is described by $I^{\pi}=1/2^{-}$, which would be
the lowest negative-parity state. Factually, among the observed
$\beta$-decay transitions from $^{17}$B, the strongest transition
$I_{\beta}=33(4)$\% is the branch to the $E_{\rm x}=2.71(2)$~MeV
state in $^{17}$C, to which $I^{\pi}=1/2^{-}$ has been assigned in
the present $\chi_{\nu}^{2}$ analysis.  The relative $B{\rm (GT)}$
strengths of the $^{17}$B $\beta$ decay are also shown in
Fig.~\ref{F:SHELLMODEL} in a similar manner as for $^{13}$C and
$^{15}$C.

%%
%% 3/2- state
%%
The decay branch to the $3/2_{1}^{-}$ state is the second strongest
transition in both the $\beta$ decays of $^{13}$B and $^{15}$B,
although the ordering of $3/2_{1}^{-}$ and $5/2_{1}^{-}$ differs for
$^{13}$C and $^{15}$C due to the appearance of the $p$-$sd$
cross-shell interactions in $^{15}$C. These transitions are
considered to be related to the single particle GT transition
${\nu}{p_{3/2}}{\rightarrow}{\pi}{p_{3/2}}$, which results in the
${\nu}{p_{3/2}}$-hole configurations in the daughter nuclei $^{13}$C
and $^{15}$C. 
%%--------new sentense------> 
Similary to the above $1/2_{1}^{-}$ case,
the ${\nu}{p_{3/2}}{\rightarrow}{\pi}{p_{3/2}}$ GT transition can
uniquely populate the single ${\nu}{p_{3/2}}$-hole state, suggesting 
its large transition strength.
%%<--------new sentense------
The same single-particle GT transition can also occur
in the $^{17}$B $\beta$ decay. The gap energy between the
${\nu}{p_{3/2}}$- and ${\nu}{p_{1/2}}$-hole states should be
comparable in both $^{15}$C and $^{17}$C, provided the spin-orbit
splitting remains relatively unchanged. From this viewpoint,
$I^{\pi}=3/2^{-}$ can be assigned to the 3.93(2)-MeV state in
$^{17}$C, because the energy difference $\Delta E= 1.22$~MeV between
the 3.93(2)-MeV state and ${1/2}^{-}$ 2.71(2)-MeV state in $^{17}$C
is comparable to $\Delta E=$ 1.55~MeV, which is the difference
between the $1/2_{1}^{-}$ and $3/2_{1}^{-}$ states in $^{15}$C, and
also because the second strongest $B{\rm (GT)}$ value was obtained
for the transition to the 3.93(2)-MeV state. It should be noted that 
this $I^{\pi}$ assignment agrees with the $I^{\pi}=(3,\ 5)/2^{-}$
assignment in the present $\chi_{\nu}^{2}$ analysis. 

%%
%% 5/2- state
%%
We may assign $5/2^{-}$ to the 4.05(2)-MeV state, which is a
neighboring partner state to the $3/2_{1}^{-}$ 3.93(2)-MeV state, by
taking into account the similarity of the level structure in this
case to that of $^{15}$C, although we cannot disregard the fact that
there are three possible candidates $I^{\pi}=(1,\ 3,\ 5)/2^{-}$ in
the $\chi_{\nu}^{2}$ analysis. The obtained ${\log}{ft}=6.0(1)$ value
for the transition to the 4.05(2)-MeV state is not sufficiently small
to be definitely assigned to the GT transition. However, ${\log}{ft}$
values to the corresponding $5/2_{1}^{-}$ states at 7.547 and
4.220~MeV in $^{13}$C and $^{15}$C are also as large as ${\log}{ft}=$ 
5.33(10)~\cite{AJ91} and 5.09(9)~\cite{HAR99_15B}, respectively,
indicating the similarity among their transitions strengths. 
%%---------------------------------------------------------
%% 5/2- state は psd model スペース内では、単一の 
%% 
As investigated in the above,
the GT transition ${\nu}{p_{3/2}({\nu}p_{1/2})}{\rightarrow}{\pi}{p_{3/2}}$
uniquely populate the $1/2_{1}^{-}(3/2_{1}^{-})$ state having 
single ${\nu}{p_{1/2}}$-particle (${\nu}{p_{3/2}}$-hole) state
properties, which causes the concentration of transition strengths.
However, the $5/2_{1}^{-}$ state cannot be formed by an unpaired
particle or hole in the $p$-$sd$ shell unlike the $1/2_{1}^{-}$ and
$3/2_{1}^{-}$ states. It should be a mixed configuration such as
$|^{17}{\rm C}{\rangle}^{5/2^{-}}=$
$|(({\pi}p_{1/2})_{1}({\pi}p_{3/2})_{-1})^{1^{+},\ 2^{+}}\otimes({\nu}{p_{3/2}})_{-1}({\nu}sd)_{4}^{0^{+}}{\rangle}^{5/2^{-}}$,
which prevents the concentration of the transition strength.
Moreover, the GT transition
${\nu}{p_{3/2}({\nu}p_{1/2}}){\rightarrow}{\pi}{p_{1/2}}$, which 
can populate these configurations, can also populate
the states up to $I^{\pi}=7/2^{-}(5/2^{-})$, further fragments the
transition strength.
Thus, the $5/2^{-}$ assignment to the 4.05(2)-MeV state seems
reasonable, although further experimental observations are necessary
for a definite assignment.

\begin{figure}[bthp]
\begin{center}
 \includegraphics[width=.80\textwidth]{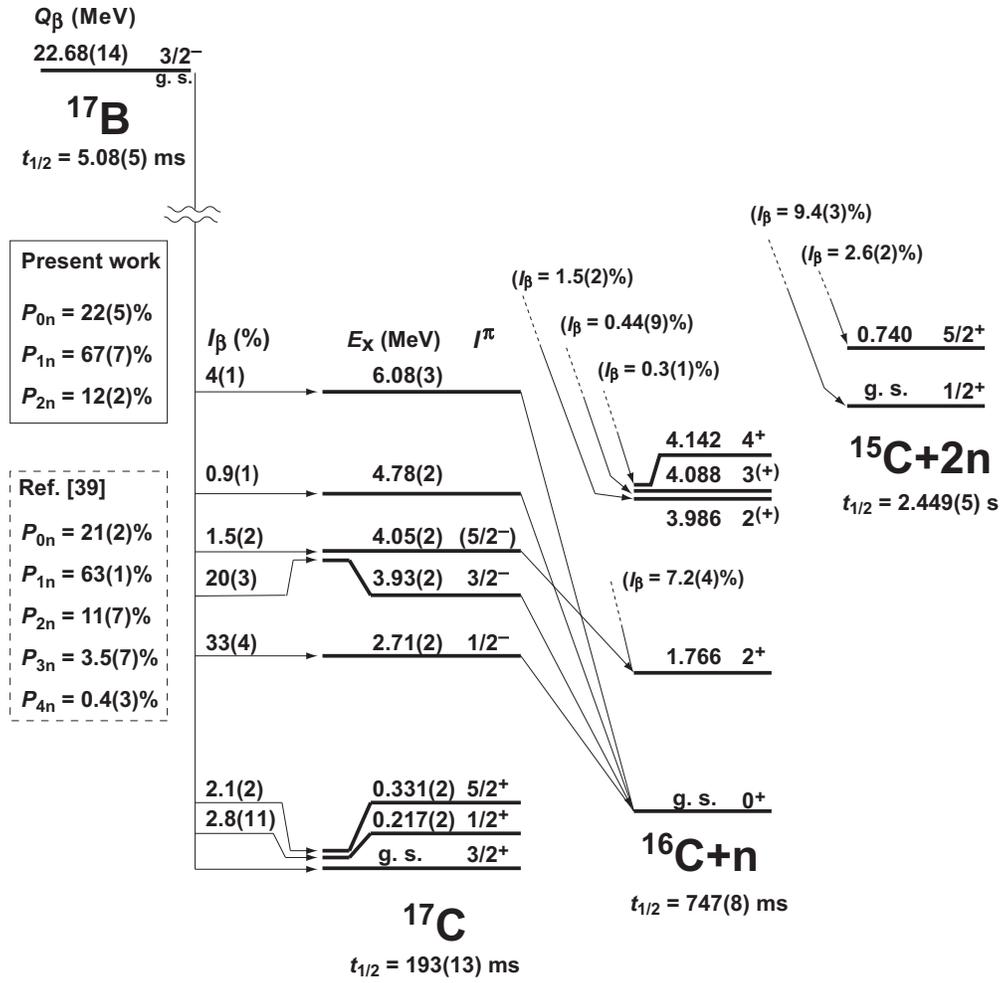}
\caption{Decay scheme of $^{17}$B constructed in the present
  work. Unassigned levels $E_{\rm x}=$ 4.78(2) and 6.08(3)~MeV are
  also shown.
\label{F:DECAY}}
  \end{center}
\end{figure}

%%---------------------------------------------------------
Including the above mentioned $I^{\pi}$ assignment, results obtained
in the present work for the $^{17}$B $\beta$ decay are shown in
Fig.~\ref{F:DECAY} as the decay scheme. 

%%------------------------------------------------------------------70
%% FIG13 should be planced around here
%%------------------------------------------------------------------70
%  \begin{figure}[bthp]
%  \begin{center}
%   \includegraphics[width=.80\textwidth]{FIG13_UENO_DCYRES.eps}
%  \caption{Decay scheme of $^{17}$B constructed in the present
%    work. Unassigned levels $E_{\rm x}=$ 4.78(2) and 6.08(3)~MeV are 
%    also shown.
%  \label{F:DECAY}}
%    \end{center}
%  \end{figure}
%%------------------------------------------------------------------70

\subsection{Comparison with shell-model calculations}

\begin{figure}[tbhp]
\begin{center}
 \includegraphics[width=.80\textwidth]{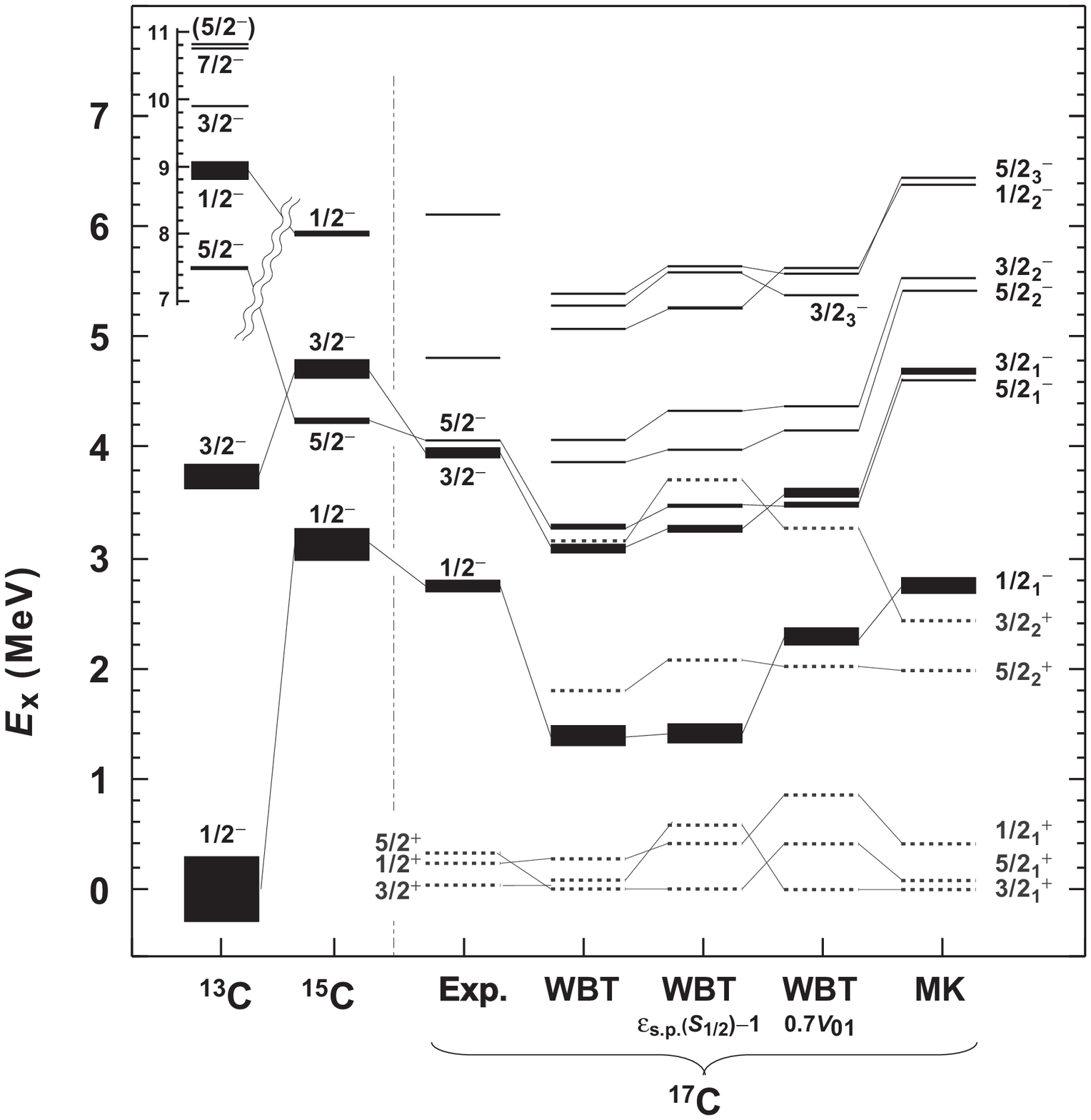}
\caption{Comparison of the obtained $\beta$-decay feeding excited
  states in $^{17}$C with shell-model calculations, with positive and
  negative parity states indicated by dashed and solid lines,
  respectively. The negative-parity states in $^{13}$C and $^{15}$C
  are also shown.
\label{F:SHELLMODEL}}
\end{center}
\end{figure}

In Fig.~\ref{F:SHELLMODEL}, the data of the excited states of
$^{17}$C observed in the present study are compared with the results
of shell-model calculations~\cite{OXBASH} with two different sets of
effective interactions, PSDWBT~\cite{WBP} and PSDMK~\cite{MK},
denoted by WBT and MK, respectively.

With respect to the low-lying positive-parity states, the GS has been
experimentally assigned to 
$I^{\pi}=3/2^{+}$~\cite{BAU98,BAZ98,SAU00,MAD01,SAT08,SCH95,OGW02_17C},
and  the 210-keV and 331-keV states have been assigned to $1/2^+$ and
$5/2^+$~\cite{ELE05,KON09_17C,SUZ08_17C}. Here, we assumed that the
excited state at $E_{\rm x} \sim 210$~keV and the observed 217(2)-keV
state are identical. The shell-model calculations predict the
existence of three positive-parity states, $I^{\pi} = 3/2^{+}$,
$1/2^+$, and $5/2^+$, below the neutron separation energy $S_{\rm n}
= 0.729(18)$~MeV. However, their ordering changes, depending on the
choice of effective interactions. For the $I^{\pi}$ value of the GS,
the MK interaction predicts the $3/2^{+}$ correctly, whereas the WBT
interaction predicts $5/2^{+}$. The experimental and theoretical
positive-parity states in $^{17}$C are indicated by dashed lines in
Fig.~\ref{F:SHELLMODEL}. We note that the experimentally determined
ordering of $1/2^{+}$ and $5/2^{+}$ is inverted both in the MK and
WBT calculations.

Next, the observed negative-parity states were compared. As expected
from the above comparison for positive-parity states, the $1/2^{-}$
state is predicted as the lowest negative-parity state in $^{17}$C
according to both the WBT and MK interactions, which thus supports
the assignment of $1/2^{-}$ to the $E_{\rm x}=2.71(2)$~MeV state. The
$E_{\rm x}$ energy for this state, predicted using the MK
interaction, agrees well with the observation, while it is
$\sim$1~MeV lower when determined using the WBT interaction. For the
next two states at 3.93(2) and 4.05(2)~MeV, $I^{\pi} = 3/2^{-}$ and
$(5/2^{-})$ were respectively assigned herein. The energy gap of
their centroid, whose displacement is only 110~keV from the GS, is
6.0~MeV. Such a pair of neighboring states are also predicted using
both the MK and WBT interactions. The MK interaction predicts the
$5/2^{-}$ and $3/2^{-}$ states at $E_{\rm x}=$ 4.62 and 4.69~MeV,
respectively, for which the centroid is $\sim$0.7~MeV higher than in
the experimental observation. In contrast, the ordering is inverted
in the WBT calculation, where the $3/2^{-}$ and $5/2^{-}$ states are
predicted at $E_{\rm x}=$ 3.09 and 3.26~MeV, respectively. In this
case, the centroid is lower by $\sim$0.8~MeV in the direction
opposite to that observed in the MK interaction. If we calculate
$E_{\rm x}$ for them using the $3/2_{1}^{+}$ state instead of the
predicted $5/2_{1}^{+}$ GS, the difference is found to be
$\sim$0.9~MeV. Thus, the WBT interaction based calculation
systematically predicts $\sim$1~MeV lower $E_{\rm x}$ energies for
all the low-lying negative-parity states.

In the light mass region, the phenomena of reduction in the single
particle energy of the  $s_{1/2}$ state are known to occur for
$N=7$~\cite{TAL60} and $N=9$~\cite{SUZ94_S12N9} nuclei. However, the
$\sim$1~MeV discrepancy due to the tendency of the energies of
negative-parity states in $^{17}$C to be calculated as lower with the
WBT interaction cannot be corrected by lowering the $s_{1/2}$ single
particle energy. The $^{17}$C energy levels calculated by reducing
the original $s_{1/2}$  single particle energy 
$\epsilon_{\rm s.~p.}(s_{1/2})$ to
$\epsilon_{\rm{s.~p.}}(s_{1/2})-1$~MeV are also shown in
Fig.~\ref{F:SHELLMODEL}. This modification changes the energies
$E_{\rm x}$ of the positive-parity states, and in particular, the
ordering of the low-lying states, while it does not affect the
energies $E_{\rm x}$ of the negative-parity states. In fact, the
energy displacement from the $3/2_{1}^{+}$ state is further lowered,
and hence, the discrepancy from the experimental observation is
rather exacerbated.

A further investigation was performed with the WBT interaction. In
this study, the WBT calculation was performed, with the $|V_{01}|$
values reduced by 30\%, for the $sd$ neutrons, and herein, $|V_{01}|$
denotes a diagonal matrix element of a two-body effective interaction
in a channel with angular momentum $I=0$ and isospin $T=1$ for
particles within the $sd$ orbits. The  diminished pairing energy is
attributed to the following three reasons. First, systematic
over-binding occurs for the even mass-number C isotopes $^{16}$C,
$^{18}$C, and $^{20}$C with the WBP interaction~\cite{WBP}, as
reported in the study of binding energies for the neutron-rich C
isotopes~\cite{BAZ95}. Second, the weakening of the coupling between
excess neutrons and the core in nuclei away from the stability line
causes pairing energies to be modified. A significant amount of
theoretical $V_{01}$ values of the two neutrons in the $sd$ shell are
originated from the renormalization of the two-body interaction due
to the polarization of the core~\cite{GEBROWN}. Finally, it was shown
in the study of the GS magnetic-moment measurement of
$^{17}$B~\cite{HU17B} that the use of the  pairing energy $|V_{01}|$
reduced by 30\% improves agreements of the experimental and
theoretical magnetic moments as well as low-lying state energies of
neighboring neutron-rich nuclei. In the state structure calculated
with the $0.7|V_{01}|$  values, shown in Fig.~\ref{F:SHELLMODEL}, the
energy gap between the GS and the $1/2_{1}^{-}$ state becomes as wide
as in the experimental observation. Furthermore, the GS described by
$I^{\pi}=3/2^{+}$ was correctly reproduced. The ordering of the other
low-lying positive-parity states $1/2_{1}^{+}$ and $5/2_{1}^{+}$ is,
however, inverted from the experimental observation in the
calculations with both WBT ($0.7|V_{01}|$) and MK interactions. 

%%------------------------------------------------------------------70
%% FIG14 should be planced around here
%%------------------------------------------------------------------70
%  \begin{figure}[tbhp]
%  \begin{center}
%   \includegraphics[width=.80\textwidth]{FIG14_UENO_SM.eps}
%   \caption{Comparison of the obtained $\beta$-decay feeding excited 
%    states in $^{17}$C with shell-model calculations, with positive
%    and negative parity states indicated by dashed and solid lines, 
%    respectively. The negative-parity states in $^{13}$C and
%    $^{15}$C are also shown.
%  \label{F:SHELLMODEL}}
%  \end{center}
%  \end{figure}
%%------------------------------------------------------------------70

%%%%%%%%%%%%%%%%%%%%%%%%%%%%%%%%%%%%%%%%%%%%%%%%%%%%%%%%%%%%%%%%%%%%70
%%
%%
\section{Summary}
%%
%%
%%%%%%%%%%%%%%%%%%%%%%%%%%%%%%%%%%%%%%%%%%%%%%%%%%%%%%%%%%%%%%%%%%%%70

By virture of the large $Q_{\beta}$ window, excited states in
$^{17}$C were effectively investigated through the measurement of
$\beta$-n and $\gamma$ rays emitted in the $\beta$ decay of
$^{17}$B. In the measurement, three negative-parity states and two
inconclusive states, were identified above the neutron threshold
energy in the 1n channel of the $^{17}$B $\beta$ decay. Further, two
positive-parity states below the threshold were also observed. For
the transitions, experimental $I_{\beta}$ values were determined. We
note that in the 1n-channel, the 1767(6)-keV $\gamma$ ray from the
first excited state of $^{16}$C was observed in coincidence with the
emitted $\beta$-n, which changes the reported $\beta$-decay scheme of
$^{17}$B and level structure of $^{17}$C. Apart from the 1767(6)-keV
lines, several de-excitation $\gamma$ lines connected after the
$\beta$-n emission were identified. In the present work, the
$\beta$-NMR technique was combined with the $\beta$-delayed particle
measurements using a fragmentation-induced spin-polarized $^{17}$B
beam. This new scheme allowed us to determine the spin parity of
$\beta$-decay feeding excited states based on the difference in the
$\beta$-decay asymmetry parameter, which took three discrete values
depending on the final state spin for a common initial spin, if the
states are connected through the GT transition. As a result, $I^{\pi}
= 1/2^{-}$, $3/2^{-}$, and ($5/2^{-}$) have been assigned to the
observed states at $E_{\rm x}=$ 2.71(2), 3.93(2), and 4.05(2)~MeV in
$^{17}$C, respectively. The observed gap energy between low-lying
positive and negative-parity states is 1$\sim$2~MeV larger than that
predicted by the shell-model calculation with the WBT
interaction. This discrepancy can be reduced by assuming that the
pairing energy for neutrons in the $sd$ shell of a neutron-rich
nucleus diminishes by about 30\%, although it cannot be resolved by
reducing the $s_{1/2}$ single-particle energy.

%%%%%%%%%%%%%%%%%%%%%%%%%%%%%%%%%%%%%%%%%%%%%%%%%%%%%%%%%%%%%%%%%%%%70
%%
%%
%% Acknowledgements
%%
%%
%%%%%%%%%%%%%%%%%%%%%%%%%%%%%%%%%%%%%%%%%%%%%%%%%%%%%%%%%%%%%%%%%%%%70

\section*{Acknowledgments}
The authors are grateful to the staffs at the RIKEN Ring Cyclotron for
their support during the execution of the experiments. This work was
supported in part by a Grant-in-Aid for Scientific Research from the
Ministry of Education, Science, Sports and Culture. The experiment
was performed at RIKEN under the Experimental Program R206n(5B). 

%%%%%%%%%%%%%%%%%%%%%%%%%%%%%%%%%%%%%%%%%%%%%%%%%%%%%%%%%%%%%%%%%%%%70
%%
%%
%% References
%%
%%
%%%%%%%%%%%%%%%%%%%%%%%%%%%%%%%%%%%%%%%%%%%%%%%%%%%%%%%%%%%%%%%%%%%%70
%% 

\newpage

%%------------------------------------------------------------------70
%% Figure-1
%%------------------------------------------------------------------70
%%------------------------------------------------------------------70

%%------------------------------------------------------------------70
%% Figure-2
%%------------------------------------------------------------------70
%%------------------------------------------------------------------70

%%------------------------------------------------------------------70
%% Figure-3
%%------------------------------------------------------------------70
%%------------------------------------------------------------------70

%%------------------------------------------------------------------70
%% Figure-4
%%------------------------------------------------------------------70
%%------------------------------------------------------------------70

%%------------------------------------------------------------------70
%% Figure-5
%%------------------------------------------------------------------70
%%------------------------------------------------------------------70

%%------------------------------------------------------------------70
%% Figure-6
%%------------------------------------------------------------------70
%%------------------------------------------------------------------70

%%------------------------------------------------------------------70
%% Figure-7
%%------------------------------------------------------------------70
%%------------------------------------------------------------------70

%%------------------------------------------------------------------70
%% Figure-8
%%------------------------------------------------------------------70
%%------------------------------------------------------------------70

%%------------------------------------------------------------------70
%% Figure-9
%%------------------------------------------------------------------70
%%------------------------------------------------------------------70

%%------------------------------------------------------------------70
%% Figure-10
%%------------------------------------------------------------------70
%%------------------------------------------------------------------70

%%------------------------------------------------------------------70
%% Figure-11
%%------------------------------------------------------------------70
%%------------------------------------------------------------------70

%%------------------------------------------------------------------70
%% Figure-12
%%------------------------------------------------------------------70
%%------------------------------------------------------------------70

%%------------------------------------------------------------------70
%% Figure-13
%%------------------------------------------------------------------70
%%------------------------------------------------------------------70

%%------------------------------------------------------------------70
%% Figure-14
%%------------------------------------------------------------------70
%%------------------------------------------------------------------70

\newpage

%%------------------------------------------------------------------70
%% TABLE-1
%%------------------------------------------------------------------70
\begin{table}[tbhp]
\catcode`?=\active \def?{\phantom{(}}
\caption{\label{T:LOWLYING_GAMMA}
Energies of the $\gamma$-ray observed as a peak in the
$\beta$-$\gamma$ coincidence measurements with $^{17}$B and $^{17}$N
gbeams below the  $^{17}$C neutron-threshold energy $S_{\rm n}=$
0.729(18)~MeV, obtained using the Ge detector, together with those
observed without a beam. Circles are used to indicate that they were
observed in each measurement. Possible $\gamma$-ray origins are
listed together in the last column, provided their energies agree
with the experimental values.} 
\begin{ruledtabular}
\begin{tabular}{cccclc}
%------------------
$E_{\gamma}$ &   no  & ?$^{17}$N? & ??$^{17}$B???
  & \multicolumn{2}{c}{possible background}\\
(keV)  &  ?beam? & $\beta$-$\gamma$ & $\beta$-$\gamma$
  & \ \ \ \ \ sources & (keV) \\
%------------------
\hline\\[-3mm]
%------------------
?67(1) & $\times$ & $\circ$ & $\circ$ &
	$\begin{cases}
		{\rm KX}_{\alpha 1}{\rm (Pt)}\\
		^{\rm 73m}{\rm Ge}
	\end{cases}$ &
	\begin{tabular}{c}
		66.8\\ 
		66.6
	\end{tabular}\\  \\[-3mm]
%------------------
?77(2) & $\circ$ & $\circ$ & $\circ$   &
	$\begin{cases}
		{\rm KX}_{\alpha 1}{\rm (Pb)}\\
		{\rm KX}_{\beta 1}{\rm (Pt)}\\
		{\rm KX}_{\alpha 1}{\rm (Bi)}
	\end{cases}$ &
	\begin{tabular}{c}
		75.0\\
		75.7\\
		77.1
	\end{tabular}\\  \\[-3mm]
%------------------
 ?88(2) & $\times$& $\times$& $\circ$ 
        & ??${\rm KX}_{\beta 1}$(Pb) & 87.3 \\
%------------------
217(2)  & $\times$& $\times$& $\circ$
        & ?($^{228}$Ac)	& (216.0) \\
%------------------
242(1)  & $\circ$ & $\times$& $\circ$ & 
	$\begin{cases}
		^{212}{\rm Pb}\\
		^{224}{\rm Ra}\\
		^{214}{\rm Pb}
	\end{cases}$ &
	\begin{tabular}{c}
		238.6\\
		241.0\\
		242.0
	\end{tabular}\\  \\[-3mm]
%------------------
295(2)  & $\times$ & $\times$& $\circ$
        & ?($^{214}$Pb)	& (295.2) \\
%------------------
331(2)  & $\times$& $\times$& $\circ$   
        & ?($^{228}$Ac)	& (328.0) \\
%------------------
352(1)  & $\circ$ & $\circ$ & $\circ$   
        & ??$^{214}$Pb	& 351.9	\\
%------------------
511(2)  & $\circ$ & $\circ$ & $\circ$   
        & ??annihilation	& 511.0	\\
%------------------
596(5)  & $\times$& $\circ$ & $\circ$   & 
	$\begin{cases}
		^{73}{\rm Ge}({\rm n},\gamma)\\
		^{74}{\rm Ge}({\rm n},{\rm n}$'$\gamma)
	\end{cases}$ &
	596.4	\\ \\[-3mm]
%------------------
609(2)  & $\circ$ & $\circ$ & $\circ$   
        & ??$^{214}$Bi	& 609.3	\\
%------------------
696(8)  & $\times$& $\circ$ & $\circ$   
        & ??$^{72}$Ge(n,n'$\gamma$) & 693 \\
%------------------
\end{tabular}
\end{ruledtabular}
\end{table}
%%------------------------------------------------------------------70

%%------------------------------------------------------------------70
%% TABLE-2
%%------------------------------------------------------------------70
\begin{table*}[bthp]
\catcode`?=\active \def?{\phantom{1}}
\caption{\label{T:0N_BRANCHES}
Properties of $\beta$-$\gamma$ rays observed in the $^{17}$B $\beta$
decay. This table lists the $\gamma$ transition energies
($E_\gamma$), decay modes  classified with the multiplicity of
$\beta$-n emission, $\gamma$-ray emitter, level energies 
(${E_{\rm x}}^{\gamma}$) in the $\gamma$ emitter, level energies
($E_{\rm x}$($^{17}$C)) in $^{17}$C, $\gamma$- and $\beta$-ray
intensities per $^{17}$B decay ($I_{\gamma}$ and $I_{\beta}$), and
${\log}{ft}$ values. For obtaining the $I_{\beta}$ value at
$E_{\gamma}=1767(6)$~keV, the intensities of the cascade decay from
2212(10), 2322(6), and 2379(7)~keV were subtracted.}
\begin{ruledtabular}
\begin{tabular}{cccccccc}
% %%
% \hline\hline
\mbox{\begin{tabular}{c}$E_{\gamma}$\\(keV)\end{tabular}} &
\mbox{\begin{tabular}{c}decay\\mode\end{tabular}} &
\mbox{\begin{tabular}{c}$\gamma$\\emitter\end{tabular}} &
\mbox{\begin{tabular}{c}
${E_{\rm x}}^{\gamma}$\\(keV)\end{tabular}} &
\mbox{\begin{tabular}{c}
$E_{\rm x}$($^{17}$C)\\(keV)\end{tabular}} &
\mbox{$I_{\gamma}$ (\%)} &
\mbox{$I_{\beta}$ (\%)} &
\mbox{${\log}{ft}$} \\
\hline
%%-----------Eg=217(2)
?217(2)? & $\beta$(0n) & $^{17}$C 
       &  ?217(2)????? &  217(2) & 2.8(11) & 2.8(11) & 6.1(2) \\
?331(2)? & $\beta$(0n) & $^{17}$C 
       &  ?331(2)????? &  331(2) & 2.1(2)? & 2.1(2)? & 6.22(5) \\
?740(2)?  & $\beta$(2n) & $^{15}$C 
       &  ?740.0(15)???? & not identified & 2.6(2)? & 2.6(2)? \\
1767(6)? & $\beta$(1n) & $^{16}$C 
       & 1766(10)???? & 4050(20) & 9.4(3)? & 7.2(4)? & 5.29(3) \\
1382(5)? & $\beta$(0n)-$\beta$($^{17}$C) & $^{17}$N 
       & 1373.8(3)??? & $-$      & 5.9(6)? & $-$ \\
1855(8)? & $\beta$(0n)-$\beta$($^{17}$C) & $^{17}$N 
       & 1849.5(3)??? & $-$      & 4.5(6)? & $-$ \\
2212(10) & $\beta$(1n) & $^{16}$C 
       & 3986(7)???? & not identified & 1.5(2)? &  1.5(2)?  &  \\
2322(6)? & $\beta$(1n) & $^{16}$C 
       & 4088(7)???? & not identified & 0.44(9) &  0.44(9)  &  \\
2379(7)? & $\beta$(1n) & $^{16}$C 
       & 4142(7)???? & not identified & 0.3(1)? &  0.3(1)?  &  \\
5290(12) & $\beta$(2n)-$\beta$($^{15}$C) & $^{15}$N 
       & 5298.822(14) & $-$ & 7.5(11) & $-$ &  \\
%%-----------
\end{tabular}
\end{ruledtabular}
\end{table*}
%%------------------------------------------------------------------70

%%------------------------------------------------------------------70
%% TABLE-3
%%------------------------------------------------------------------70
\begin{table*}[bthp]
\catcode`?=\active \def?{\phantom{0}}
\caption{\label{T:1N_BRANCHES}
Levels in $^{17}$C observed in the 1n-decay channel following the $\beta$
decay of $^{17}$B, which are listed along
with the excitation energies $E_{\rm x}$ for $^{17}$C and the neutron-kinetic
energies $E_{\rm n}$, from which the $E_{\rm x}$ values are converted.
The inconclusive levels are indicated by $\{$ $\}$.
Determined transition strengths $I_{\beta}$ and values of ${\log}{ft}$
for the $^{17}$C levels are also listed. For details, see the text.}
\begin{ruledtabular}
\begin{tabular}{rccccccccccc}
\phantom{$\{$}
& \mbox{\begin{tabular}{c} $E_{\rm n}$\\(MeV) \end{tabular}}
& \mbox{\begin{tabular}{c} $E_{\rm x}$\\(MeV) \end{tabular}}
& \mbox{\begin{tabular}{c} width\\(MeV)       \end{tabular}}
& \mbox{\begin{tabular}{c} $I_{\beta}$\\(\%)  \end{tabular}}
& \mbox{${\log}{ft}$}
& \mbox{$B$(GT)}
& \mbox{$A_{\beta}P$ (\%)}
& \mbox{$A_{\beta}$}
& \mbox{\begin{tabular}{c} $2I^{\pi}$\\($\chi_{\nu}^{2}$ analysis) 
        \end{tabular}}
& \mbox{\begin{tabular}{c} $2I^{\pi}$\\(end result) 
        \end{tabular}}
\phantom{$\{$}\\
\hline
%%-----------Ex=6.08 Ib=4.18(57) logft=5.28(7) BGT=0.021(3)
$\{$ & 5.04(2) &  6.08(3) & 2.5(7) & ?4(1) & 5.3(1) & 0.021(3)
	& & & & & $\}$ \\
%%-----------Ex=4.78 Ib=0.88(12) logft=6.12(7) BGT=0.0030(5)
$\{$ & 3.81(1) & 4.78(2) & 0.3(3) & ???0.9(1) & 6.1(1) & 0.003(1)
	& & & & & $\}$ \\
%%-----------Ex=4.05 Ib=1.50(20) logft=5.97(6) BGT=0.0042(6)
     & 1.46(1) & 4.05(2) & ?0.06(6) & ????1.5(2)? & 6.0(1) & 0.004(1)
	& $+$6(23)? & $-$4(15)?
	& (1$^{-}$, 3$^{-}$, 5$^{-}$) & (5$^{-}$) & \\	%% Ipi
%%-----------Ex=3.93 Ib=19.9(2.7) logft=4.86(7) BGT=0.054(9)
     & 3.01(1) & 3.93(2) & ?0.16(4) & 20(3) & 4.9(1) &  0.05(1)
	& ?$-0.1(15)$ & ??$+$0.04(99)
	& (3$^{-}$, 5$^{-}$) & 3$^{-}$ & \\
%%-----------Ex=2.71 Ib=32.6(4.4) logft=4.78(6) BGT=0.065(9)
     & 1.86(1) & 2.71(2) & ?0.04(1) & 33(4) & 4.8(1) & 0.07(1)
	& $+$1.6(8)? & $-$1.0(5)
	& $1^{-}$ & $1^{-}$ & \\
%%-----------
\end{tabular}
\end{ruledtabular}
\end{table*}
%%------------------------------------------------------------------70

\end{document}